\documentclass[12pt, a4paper]{article}
\pdfoutput=1
\usepackage{graphicx}
\usepackage{amsmath, amssymb}
\usepackage{bm}
\usepackage{slashed}
\usepackage{cite}
\usepackage{subfigure}

\newcommand{\Slash}[1]{{\ooalign{\hfil/\hfil\crcr$#1$}}}

\usepackage{hyperref}

\allowdisplaybreaks[1]

\begin{document}

\begin{titlepage}

\def\thefootnote{\fnsymbol{footnote}}

\begin{center}

\hfill IPMU~15-0038 \\
\hfill KANAZAWA-15-02\\
\hfill FTPI-MINN-15/16

\vspace{2cm}
{\large\bf QCD Effects on Direct Detection of  
 Wino Dark Matter}

\vspace{1cm}
 {\large Junji Hisano}$^{\it (a,b,c)}$, 
 {\large Koji Ishiwata}$^{\it (d)}$, 
 {\large Natsumi Nagata}$^{\it (c,e)}$

 \vspace{1cm}

 {\it $^{(a)}${Kobayashi-Maskawa Institute for the Origin of Particles and the Universe, \\ Nagoya University, Nagoya 464-8602, Japan}}\\

 {\it $^{(b)}${Department of Physics, 
 Nagoya University, Nagoya 464-8602, Japan}}

 \vspace{0.2cm}
 {\it $^{(c)}$Kavli IPMU (WPI), University of Tokyo, Kashiwa, Chiba
 277-8584, Japan} 

 \vspace{0.2cm}

% {\it $^{(d)}$ Deutsches Elektronen Synchrotron DESY, Notkestrasse 85,
%          22607 Hamburg, Germany}\\

 {\it $^{(d)}$ 
Institute for Theoretical Physics, Kanazawa University, Kanazawa 920-1192, Japan}\\

 \vspace{0.2cm}

{\it $^{(e)}$William I. Fine Theoretical Physics Institute, School of
 Physics and Astronomy, University of Minnesota, Minneapolis, MN 55455,
 USA}\\

 \vspace{1cm}

 \abstract{We complete the calculation of the wino-nucleon scattering
 cross section up to the next-to-leading order in $\alpha_s$. We assume
 that the other sparticles are decoupled and wino interacts with the
 Standard Model particles via the weak interaction. As a result, the
 uncertainties coming from the perturbative QCD are significantly
 reduced to be smaller than those from the nucleon matrix elements. The
 resultant scattering cross section is found to be larger than the
 leading-order one by about 70\%, which is well above the neutrino
 background. In the limit of large wino mass the spin-independent
 scattering cross section with proton turns out $\sigma_{\text{SI}}^p =
 2.3~{}^{+0.2}_{-0.3}~{}^{+0.5}_{-0.4}\times 10^{-47}~ \text{cm}^2$
 (errors come from perturbative calculation and input parameters,
 respectively).  The computation for a generic SU(2)$_L$ multiplet dark
 matter is also presented.  }

 \end{center}
 \end{titlepage}

\renewcommand{\theequation}{\thesection.\arabic{equation}}
\renewcommand{\thepage}{\arabic{page}}
\setcounter{page}{1}
\renewcommand{\thefootnote}{\#\arabic{footnote}}
\setcounter{footnote}{0}

%%%%%%%%%%%%%%%%%%%%%%%%%%%%%%%%%%%%%%%%%%%%%%%%%%%%%%%%
\section{Introduction}
\label{sec:intro}
\setcounter{equation}{0} 
%%%%%%%%%%%%%%%%%%%%%%%%%%%%%%%%%%%%%%%%%%%%%%%%%%%%%%%

Weakly-interacting massive particles (WIMPs) are promising candidates
for dark matter (DM) in the Universe.  Many theoretical models beyond
the Standard Model predict WIMPs and it is known that the thermal WIMP
scenario can explain the present energy density of dark matter in
those models.  The early stage of the experiments at the Large Hadron
Collider, however, has found no evidence for new physics near the
electroweak scale so far. In particular, the experiments give severe
bounds on new colored particles, such as gluino and squarks in the
supersymmetric (SUSY) models \cite{Chatrchyan:2014lfa}. This situation
may imply that most of the new particles in the high energy theory
have masses much larger than the electroweak scale and only a WIMP,
probably accompanied with some non-colored particles, is accessible in
the TeV-scale experiments.

 The current experimental consequences would fit in with a simple SUSY
 breaking scenario. If the SUSY breaking is induced by non-singlet
 chiral superfields (as is often the case with the dynamical SUSY
 breaking \cite{Affleck:1983rr}), gaugino masses are induced by the
 anomaly mediation mechanism \cite{Randall:1998uk, Giudice:1998xp} and
 thus suppressed by a loop factor compared with the gravitino mass. A
 generic K\"{a}hler potential gives masses of the order of the
 gravitino mass to scalar particles and higgsino. In this framework,
 the neutral wino is found to be the lightest SUSY particle and thus
 becomes a candidate for dark matter in the Universe. Actually, its
 thermal relic abundance explains the observed energy density of DM if
 the wino DM has a mass of $2.7-3.1$~TeV \cite{Hisano:2006nn}. For
 relatively light wino DM, on the other hand, the non-thermal
 production via the late time decay of gravitino could be invoked to
 provide the correct abundance of DM \cite{Gherghetta:1999sw,
   Moroi:1999zb}. As this scenario \cite{Giudice:1998xp,Wells:2003tf}
 requires the SUSY breaking scale to be much higher than the
 electroweak scale, a relatively heavy mass for the Higgs boson is
 predicted \cite{Giudice:2011cg}, which is consistent with the
 observed value $m_h\simeq 125$~GeV \cite{Aad:2012tfa}. Such a high
 SUSY-breaking scale is phenomenologically desirable since it relaxes
 the SUSY flavor and CP problems \cite{Gabbiani:1996hi}, the
 dimension-five proton decay problem \cite{Liu:2013ula}, and some
 cosmological problems \cite{Kawasaki:2008qe}. Gauge coupling
 unification is found to be still preserved with good accuracy
 \cite{Hisano:2013cqa}. For these reasons, the wino DM scenario
 attracts a lot of attention, and its phenomenology has been studied
 widely.

A lot of efforts have been dedicated to searching for the wino DM. A
robust constraint is provided by the Large Hadron Collider experiment;
charged winos with a mass of 270~GeV or less have been excluded at
95\% C.L. \cite{Aad:2013yna}. For prospects of the wino search in
future collider experiments, see Ref.~\cite{Shingo:2013aja}. On the
other hand, signal of the wino DM may be detected in cosmic ray
observations. Since the wino DM has large annihilation cross section
\cite{Hisano:2003ec, Hisano:2004ds}, cosmic rays from annihilating
winos are promising tools to detect the wino DM indirectly.  The mass
of wino DM $M$ is constrained as $320~\text{GeV}\leq M\leq
2.25~\text{TeV}$ and $2.43~\text{TeV}\leq M \leq 2.9~\text{TeV}$ at
95\% C.L. \cite{Bhattacherjee:2014dya} by using gamma ray data from
dwarf spheroidal galaxies provided by Fermi-LAT collaboration
\cite{Ackermann:2013yva}. Gamma rays from the Galactic
center provided by the H.E.S.S.~\cite{Abramowski:2013ax} may give a
strong limit on the wino DM, though the consequences are quite
dependent on the DM density profile used in the
analysis~\cite{Cohen:2013ama}.  Developments in both theory
\cite{Baumgart:2014vma} and observation enable us to probe a wide
range of mass region of the wino DM in future indirect detection
experiments.

Direct detection of dark matter is another important experiment to
study the nature of dark matter.
%The DM direct detection experiments are also of importance to study
%the nature of DM.
Currently the most stringent limits are provided by the LUX experiment
\cite{Akerib:2013tjd}; it sets an upper limit on the spin-independent
(SI) WIMP-nucleon elastic scattering cross section as $\sigma_{\rm
  SI}<7.6\times 10^{-46}~{\rm cm}^2$ at a WIMP mass of
33~GeV. Moreover, various future projects with ton-scale detectors are
now ongoing and expected to have significantly improved
sensitivities.  To test the wino DM scenario in the direct detection
experiments, one needs to evaluate the wino-nucleon scattering cross
section precisely, with the theoretical uncertainties being
sufficiently controlled. This scattering is induced by loop diagrams
if the higgsino and squarks are much heavier than wino
\cite{Hisano:2004pv}. At present, the leading order (LO) calculation
for the scattering cross section is given in the literature
\cite{Hisano:2010fy, Hisano:2010ct, Hisano:2011cs, Hisano:2012wm}; in
these works, the SI scattering cross section with a nucleon is
evaluated as $\sigma_{\rm SI}\sim 10^{-47}~{\rm cm}^2$. For other
relevant works, see Refs.~\cite{Cao:2011ht, Hill:2011be}.  Since the
predicted scattering rate of the wino DM is larger than those of the
neutrino backgrounds \cite{Billard:2013qya}, one expects that the
future direct detection experiments may eventually catch a signal of
the wino DM. However, it was pointed out by the authors of
Ref.~\cite{Hill:2011be} that the present calculation may suffer from
large uncertainties. They further found that these uncertainties
mainly come from the neglect of the higher order contribution in
perturbation theory, not from the error of the nucleon matrix
elements, which may alter the SI cross section by a factor. To reduce
the uncertainties, therefore, we need to go beyond the LO calculation.

In this paper, we complete this calculation up to the next-to-leading
order (NLO) in the strong coupling constant $\alpha_s$. For this
purpose, we first reformulate the computation based on the effective
theoretical approach. The relevant interactions are expressed in terms of the
effective operators, whose Wilson coefficients are given up to the NLO with
respect to $\alpha_s$. The coefficients are evolved down to the scale at
which the nucleon matrix elements of the effective operators are
evaluated, by means of the renormalization group equations (RGEs). This
procedure allows us to include the NLO QCD effects systematically.

The rest of the paper is organized as follows. In the next section, we
describe the formulation mentioned above. All of the matching
conditions as well as the RGEs are presented here. Then, in
Sec.~\ref{sec:error}, we show our results for the SI scattering cross
section and discuss the uncertainties of the calculation. In
Sec.~\ref{sec:ewimp}, we also present the results for a generic
SU(2)$_L$ multiplet DM. Those who are interested in a quick reference
may find our results in these two
sections. Section~\ref{sec:conclusion} is devoted to conclusion and
discussion.

%%%%%%%%%%%%%%%%%%%%%%%%%%%%%%%%%%%
\section{Formalism}
\label{sec:formalism}
\setcounter{equation}{0} 
%%%%%%%%%%%%%%%%%%%%%%%%%%%%%%%%%%%

In this section, we give a formalism to evaluate the SI scattering
cross section of the wino DM with a nucleon. We will carry out the
calculation up to the NLO in the strong coupling constant
$\alpha_s$. The formalism given here is based on the method of
effective field theories, which consists of the following three
steps. Firstly, we obtain the effective operators at the electroweak
scale $\mu_{W}^{}\simeq m_Z^{}$ ($m_Z$ is the mass of the $Z$ boson) by
integrating out heavy particles whose masses are not less than the
electroweak scale. This step is carried out in terms of the operator
product expansions (OPEs). Secondly, we evolve the Wilson coefficients
of the effective operators using the RGEs down to the scale at which
the nucleon matrix elements of the operators are evaluated. Finally,
we express the SI effective coupling of a wino DM with a nucleon in
terms of the Wilson coefficients and the nucleon matrix elements. From
this effective coupling, one readily obtains the SI scattering cross
section.

%%%%%%%%%%%%%%%%%%%%%%%%%%%%%%%%%%%%
\subsection{Effective Lagrangian}
%%%%%%%%%%%%%%%%%%%%%%%%%%%%%%%%%%%

First let us formulate the effective Lagrangian which gives rise to
the SI interactions of the wino DM with quarks and gluon. The
effective Lagrangian comprises two types of the higher dimension
operators---the scalar and the twist-2 type operators---as follows
\cite{Drees:1993bu}:
\begin{align}
 {\cal L}_{\rm eff}
&=
\sum_{i=q,G}C^i_{\rm S} {\cal O}^i_{\rm S}
+\sum_{i=q,G}
(C^i_{\rm T_1} {\cal O}^i_{\rm T_1}+
C^i_{\rm T_2} {\cal O}^i_{\rm T_2})~,
\end{align}
with
\begin{align}
 {\cal O}^q_{\rm S}&\equiv m_q\bar{\chi}^0\chi^0\bar{q}q~,\nonumber \\
 {\cal O}^G_{\rm S}&\equiv \frac{\alpha_s}{\pi}
\bar{\chi}^0\chi^0 G^a_{\mu\nu}G^{a\mu\nu}~,
\nonumber \\
 {\cal O}^i_{\rm T_1}&\equiv \frac{1}{M}
\bar{\chi}^0i\partial^\mu \gamma^\nu\chi^0 {\cal O}^i_{\mu\nu}~,\nonumber \\
 {\cal O}^i_{\rm T_2}&\equiv \frac{1}{M^2}
\bar{\chi}^0(i\partial^\mu)(i\partial^\nu) \chi^0 {\cal O}^i_{\mu\nu}~,
\label{eq:opdef}
\end{align}
Here ${\chi}^0$, $q$, and $G^a_{\mu\nu}$ denote the wino DM, quarks,
and the field strength tensor of gluon field, respectively; $m_q$ are
the masses of quarks; $M$ is the mass of the wino DM; ${\cal
  O}^q_{\mu\nu}$ and ${\cal O}^G_{\mu\nu}$ are the twist-2 operators
of quarks and gluon, respectively, which are defined by
\footnote{We have changed
the definition of ${\cal O}^G_{\mu\nu}$ by a factor of $-1$ from those in
Refs.~\cite{Drees:1993bu, Hisano:2010fy, Hisano:2010ct, Hisano:2011cs}.
We follows the conversion in Ref.~\cite{Buras:1979yt}.}
\begin{align}
 {\cal O}^q_{\mu\nu}&\equiv \frac{1}{2}\overline{q}i\biggl(
D_\mu^{}\gamma_\nu^{} +D_\nu^{}\gamma_\mu^{}-\frac{1}{2}g_{\mu\nu}^{}
\Slash{D}\biggr)q~,\nonumber \\
{\cal O}^G_{\mu\nu}&\equiv 
G^{a\rho}_{\mu} G^{a}_{\nu\rho}-\frac{1}{4}g_{\mu\nu}^{}
G^a_{\rho\sigma}G^{a\rho\sigma}~,
\label{eq:twist2def}
\end{align}
with $D_\mu$ the covariant derivative. These effective operators are
renormalized at the electroweak scale $\mu_{W}^{}\simeq m_Z^{}$ with
$N_f =5$ active quarks ($q=u,d,s,c,b$). The Wilson coefficients of the
operators are to be determined below. Notice that we have included the
strong coupling constant $\alpha_s/\pi$ in the definition of the gluon
scalar-type operator ${\cal O}^G_{\rm S}$ \cite{Hisano:2015bma}. We will
discuss the validity in the next subsection.

%%%%%%%%%%%%%%%%%%%%%%%%%%%%%%%%%%%%%%%%
\subsection{Nucleon matrix elements}
%%%%%%%%%%%%%%%%%%%%%%%%%%%%%%%%%%%%%%

%%%%%%%%%%%%%%%%%%%%%%%%%%%%%%%%%%%%%%%%%%%%%%%%
\begin{table}[t]
\caption{Mass fractions computed with the lattice simulations of QCD
 \cite{Young:2009zb, Oksuzian:2012rzb}.  }
\label{table:massfraction}
 \begin{center}
\begin{tabular}{ll|ll}
\hline
\hline
\multicolumn{2}{c|}{Proton}&
\multicolumn{2}{c}{Neutron}\cr
\hline 
$f^{(p)}_{T_u}$& 0.019(5)&$f^{(n)}_{T_u}$&0.013(3)\cr
$f^{(p)}_{T_d}$& 0.027(6)&$f^{(n)}_{T_d}$& 0.040(9)\cr
$f^{(p)}_{T_s}$&0.009(22)&$f^{(n)}_{T_s}$& 0.009(22) \cr
\hline
\hline
\end{tabular}
\end{center}
\end{table}
%%%%%%%%%%%%%%%%%%%%%%%%%%%%%%%%%%%%%%%%%%%%%%%%%%%%%%%%%

In order to compute the scattering cross section of the wino DM with a
nucleon, we need the nucleon matrix elements of the scalar and twist-2
type quark and gluon operators presented above. Since these two types
of the operators do not mix with each other under the renormalization
group (RG) flow, it is possible to consider these two types
separately.

For the scalar-type quark operators, we use the results from the
QCD lattice simulations. The
values of the mass fractions of a nucleon $N(=p,n)$, which are defined by
\begin{equation}
 f_{T_q}^{(N)}\equiv \langle N|m_q \bar{q}q|N\rangle/m_N ~,
\end{equation}
are shown in Table~\ref{table:massfraction}. Here $m_N$ is the nucleon
mass. They are taken from Ref.~\cite{Hisano:2012wm}, which are computed
with the recent results of the lattice QCD simulations
\cite{Young:2009zb, Oksuzian:2012rzb}. 

The nucleon matrix element of ${\cal O}^G_{\rm S}$, on the other hand,
is evaluated by means of the trace anomaly of the energy-momentum
tensor in QCD \cite{Shifman:1978zn}:
\begin{equation}
 \Theta^\mu_{~\mu}=\frac{\beta
  (\alpha_s)}{4\alpha_s}G^a_{\mu\nu}G^{a\mu\nu}_{}
  +(1-\gamma_m^{})\sum_{q}m_q\overline{q}q~.
\label{eq:traceanomaly}
\end{equation}
Here the beta-function $\beta(\alpha_s)$ and the anomalous dimension
$\gamma_m$ are defined by the following equations:
\begin{equation}
 \beta (\alpha_s)\equiv \mu \frac{d\alpha_s}{d \mu}~,
~~~~~~
\gamma_m m_q \equiv \mu \frac{d m_q}{d\mu}~,
\end{equation}
whose explicit forms will be given in Eqs.~(\ref{eq:beta_s}) and
(\ref{eq:gamma_m}), respectively.  By putting the operator
\eqref{eq:traceanomaly} between the nucleon states at rest, we obtain
\begin{equation}
 \langle N|\frac{\alpha_s}{\pi} G^a_{\mu\nu}G^{a\mu\nu}|N\rangle
=m_N\frac{4\alpha_s^2}{\pi\beta (\alpha_s; N_f=3)}
\bigl[
1-(1-\gamma_m)\sum_{q}f_{Tq}^{(N)}
\bigr]~.
\label{eq:gluonscalar}
\end{equation}
This formula is obtained with $N_f=3$ quark flavors. Notice that the
relation \eqref{eq:traceanomaly} is an operator equation and thus
scale-invariant. This is because the energy-momentum tensor is
corresponding to the current of the four momentums, which is a physical
quantity and thus not renormalized. As a consequence,
Eq.~\eqref{eq:gluonscalar} should hold at any scales. We will evaluate
the matrix element at the hadronic scale $\mu_{\rm had}\simeq 1~{\rm
GeV}$ in the following calculation. 

Since $\beta (\alpha_s)={\cal O}(\alpha_s^2)$, the r.h.s. of
Eq.~\eqref{eq:gluonscalar} have a size of ${\cal O}(m_N)$.
Namely, although we include a factor of $\alpha_s/\pi$ in the
definition of ${\cal O}^G_{\rm S}$, its nucleon matrix element is not
suppressed by the factor.  It should be also noted that the
scalar-type quark operator $m_q\bar{q}q$ is scale-invariant to all
orders in perturbation theory (in a mass-independent renormalization
scheme) and then the matrix element is independent of the scale at the
LO in $\alpha_s$.
This is another reason for our definition of ${\cal O}_{\rm S}^G$.

%%%%%%%%%%%%%%%%%%%%%%%%%%%%%%%%%%%%%%%%%%%%%%%%
\begin{table}[t]
\caption{Second moments of the PDFs of proton evaluated at $\mu
 =m_Z$. We use the CJ12 next-to-leading order PDFs given by the
 CTEQ-Jefferson Lab collaboration \cite{Owens:2012bv}. }
\label{table:2ndmoments}
 \begin{center}
\begin{tabular}{ll|ll}
\hline
\hline
$g(2)$ & 0.464(2) & &\\
$u(2)$ & 0.223(3) &$\bar{u}(2)$ & 0.036(2) \\
$d(2)$ & 0.118(3) &$\bar{d}(2)$ & 0.037(3) \\
$s(2)$ & 0.0258(4) &$\bar{s}(2)$ & 0.0258(4) \\
$c(2)$ & 0.0187(2) &$\bar{c}(2)$ & 0.0187(2) \\
$b(2)$ & 0.0117(1) &$\bar{b}(2)$ & 0.0117(1) \\
\hline
\hline
\end{tabular}
\end{center}
\end{table}
%%%%%%%%%%%%%%%%%%%%%%%%%%%%%%%%%%%%%%%%%%%%%%%%%%%%%%%%%

Finally, the nucleon matrix elements of the twist-2 operators are given
by the second moments of the parton distribution functions (PDFs):
\begin{align}
\langle N(p)\vert 
{\cal O}_{\mu\nu}^q
\vert N(p) \rangle 
&=m_N \Bigl(\frac{p_{\mu}p_{\nu}}{m_N^2}-\frac{1}{4}g_{\mu\nu}\Bigr)
(q^{(N)}(2;\mu)+\bar{q}^{(N)}(2;\mu))  ~,
\label{eq:q2}
\\
\langle N(p) \vert 
{\cal O}_{\mu\nu}^G
\vert N(p) \rangle
& = 
-m_N\Bigl(\frac{p_{\mu}p_{\nu}}{m_N^2}-\frac{1}{4}g_{\mu\nu}\Bigr) 
g^{(N)}(2;\mu)  ~.
\label{eq:G2}
\end{align}
with
\begin{align}
q^{(N)}(2;\mu) &= \int^1_0 dx~ x\ q^{(N)}(x,\mu)~,
\\
\bar{q}^{(N)}(2;\mu) &= \int^1_0 dx~ x\ \bar{q}^{(N)}(x,\mu)~,
\\
g^{(N)}(2;\mu) &= \int^1_0 dx~ x\ g^{(N)}(x,\mu)~.
\end{align}
Here $q^{(N)}(x,\mu)$, $\bar{q}^{(N)}(x,\mu)$ and $g^{(N)}(x,\mu)$ are
the PDFs of quark, antiquark and gluon in nucleon at the
scale $\mu$, respectively. Contrary to the case of the scalar matrix
elements, we have the values of the PDFs at various scales. In
Table~\ref{table:2ndmoments}, for example, we present the second
moments at the scale of $\mu =m_Z$. Here we use the CJ12
next-to-leading order PDFs given by the CTEQ-Jefferson Lab collaboration
\cite{Owens:2012bv}. It turns out that with the
definition of the gluon twist-2 tensor given in
Eq.~\eqref{eq:twist2def}, the second moment for gluon $g(2)$ is of the
same order of magnitude as those for quarks so that the r.h.s. of
Eqs.~\eqref{eq:q2} and \eqref{eq:G2} are ${\cal O}(m_N)$. This
justifies the definition \eqref{eq:opdef}, where we do not
include a factor of $\alpha_s/\pi$ in the definition of ${\cal O}^G_{\rm
  T_1}$ and ${\cal O}^G_{\rm T_2}$. Our definition for the gluon
operators (${\cal O}^G_{\rm S}$, ${\cal O}^G_{\rm T_1}$, and ${\cal
  O}^G_{\rm T_2}$) clarifies the order counting with respect to
$\alpha_s/\pi$ \cite{Hisano:2015bma}.

%%%%%%%%%%%%%%%%%%%%%%%%%%%%%%%%%%%%%%%%%%
\subsection{Wilson coefficients}
%%%%%%%%%%%%%%%%%%%%%%%%%%%%%%%%%%%

%%%%%%%%%%%%%%%%%%%%%%%%%%%%%%%%%%%%%%%%%%%%%%%%%%%%%%%%
\begin{figure}[t]
  \begin{center}
    \includegraphics[scale=0.4]{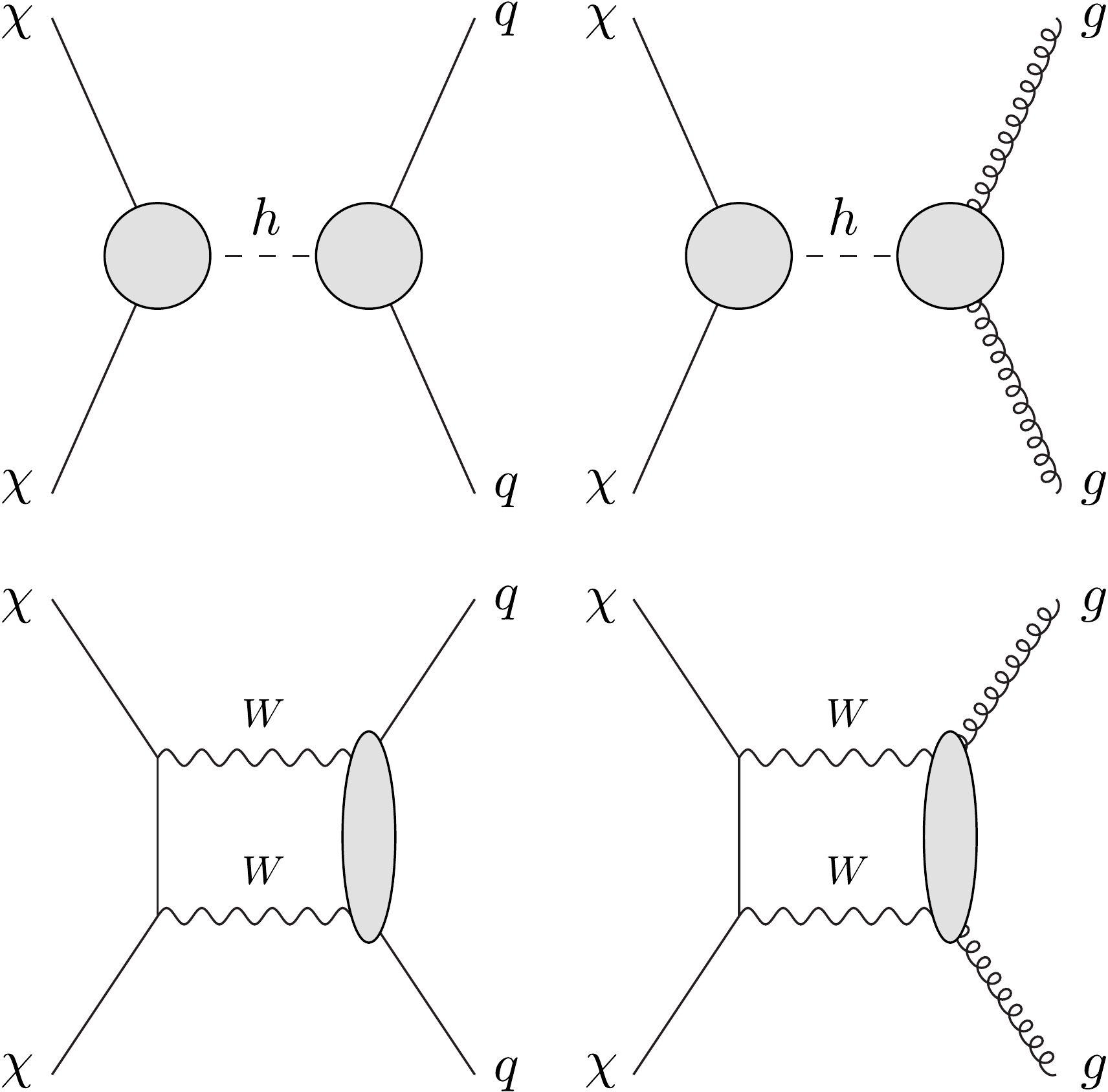}
  \end{center}
  \caption{Diagrams for wino-nucleon scattering.}
  \label{fig:DiagramTot}
\end{figure}
%%%%%%%%%%%%%%%%%%%%%%%%%%%%%%%%%%%%%%%%%%%%%%%%%%%%%%%%%

Now we evaluate the Wilson coefficients of the effective operators at
the electroweak scale $\mu_W$ to the NLO in $\alpha_s/\pi$. We
use the $\overline{\rm MS}$ scheme in the following calculation. 
The scattering of a pure neutral wino $\chi^0$ with a nucleon is induced
via the weak interactions accompanied by the charged winos
$\chi^\pm$. The interaction Lagrangian is given by
\begin{equation}
 {\cal L}_{\text{int}}= g_2\overline{\chi^0}\Slash{W}\chi^+ +\text{h.c.}~,
\end{equation}
where $g_2$ and $W_\mu$ are the SU(2)$_L$ gauge coupling constant
and the $W$ boson, respectively. Since the winos do not couple to the
Higgs field directly and the mass difference $\Delta M$ between the
neutral and charged winos is radiatively generated after the
electroweak symmetry breaking, $\Delta M$ is much smaller than the DM
mass itself or other masses which enter into our computation;
according to the recent NLO computation given in
Ref.~\cite{Ibe:2012sx}, $\Delta M\simeq 165$~MeV. Therefore, we safely
neglect it in the following discussion.
 
Before looking into the details of the calculation, we first summarize
the procedure of the computation as well as the approximations we have
used in the calculation. In Fig.~\ref{fig:DiagramTot}, we show the
diagrams which induce the couplings of wino DM with quarks and gluon,
respectively \cite{Hisano:2010fy, Hisano:2010ct, Hisano:2011cs,
  Hisano:2012wm}. These diagrams are classified into two types; one is
the Higgs exchange type like the upper two diagrams and the other is
the box diagrams corresponding to the lower two. We separately discuss
each two type.

The Higgs contribution only induces the scalar-type
operators. For the NLO-level calculation, we need to evaluate the two-
and three-loop diagrams for the quark and gluon scalar-type operators,
respectively.

For the box-type contribution, on the other hand, the NLO-level
calculation requires us to determine the Wilson coefficients of the
operators $m_q\bar{q}q$, $\frac{\alpha_s}{\pi}
G^a_{\mu\nu}G^{a\mu\nu}$, and ${\cal O}^i_{\mu\nu}$ to ${\cal
  O}(\alpha_s/\pi)$. We first carry out the OPEs of the correlation
function of the electroweak currents, as described in
Refs.~\cite{Hisano:2010ct, Hisano:2011cs}. For the scalar operators,
the NLO contribution to the OPEs of the correlation functions of
vector and axial-vector currents is evaluated in
Ref.~\cite{Broadhurst:1994qj} in the degenerate quark mass limit for
each generation. The results are directly applicable to the
contribution of the first two generations in our calculation since all
of the quarks of the generations may be regarded as
massless. Concerning the third generation contribution, the mass
difference between top and bottom quarks is significant, and thus the
mere use of the results in Ref.~\cite{Broadhurst:1994qj} is not
justified. Their contribution is, however, found to be small compared
with those of the first two generations. In our calculation, we
neglect the NLO contribution of the third generation, and take into
account the effects as a theoretical uncertainty. The Wilson
coefficients of the twist-2 operators are evaluated in
Ref.~\cite{Bardeen:1978yd} to ${\cal O}(\alpha_s/\pi)$ in the massless
limit. It is again not possible to use the results for the
contribution of the third generation, and thus we will drop the
contribution and estimate the effects as a theoretical uncertainty.
By evaluating the $W$ boson loop diagrams with this correlation
function, we then obtain the Wilson coefficients of the operators in
Eq.~\eqref{eq:opdef}.

%%%%%%%%%%%%%%%%%%%%%%%%%%%%%%%%%%%%%%%%%%%%%%%%
\begin{table}[t]
\caption{Number of loops in diagrams relevant to the ${\cal
 O}(\alpha_s/\pi)$ calculation for each operator. We also show where we
 neglect the third generation contribution at the
 NLO.  Here ``$-$'' means that there is no contribution or the
 contribution vanishes.} 
\label{table:NLOcalc}
 \begin{center}
\begin{tabular}{ll|cc|cc}
\hline
\hline
\multicolumn{2}{c|}{Operators}&
\multicolumn{2}{c|}{Higgs}&
\multicolumn{2}{c}{Box}\\
\hline
Parton&Type& LO&NLO&LO&NLO\\
\hline\hline
Quark & Scalar $C^q_{\rm S}$& 1-loop&2-loop&-&2-loop\\
(1st\&2nd)&Twist-2 $C^q_{\rm T_{1,2}}$&-&-&1-loop&2-loop\\
\hline
Quark & Scalar $C^b_{\rm S}$& 1-loop&2-loop&1-loop&2-loop (neglected)\\
($b$-quark)&Twist-2 $C^b_{\rm T_{1,2}}$&-&-&1-loop&2-loop
 (neglected)\\
\hline
Gluon & Scalar $C^G_{\rm S}$& 2-loop&3-loop&2-loop&3-loop\\
(1st \& 2nd)&Twist-2 $C^G_{\rm T_{1,2}}$&-&-&-&2-loop\\
\hline
Gluon & Scalar $C^G_{\rm S}$& 2-loop&3-loop&2-loop&3-loop (3rd gen. neglected)\\
(3rd)&Twist-2 $C^G_{\rm T_{1,2}}$&-&-&-&2-loop
 (3rd gen. neglected)\\
\hline
\hline
\end{tabular}
\end{center}
\end{table}
%%%%%%%%%%%%%%%%%%%%%%%%%%%%%%%%%%%%%%%%%%%%%%%%%%%%%%%%%

As a result, $C^q_{\rm S}$, $C^i_{\rm T_1}$,
and $C^i_{\rm T_2}$ are computed at the two-loop level, while $C^G_{\rm S}$
is evaluated at the three-loop level. In Table~\ref{table:NLOcalc}, we
summarize the number of loops in diagrams relevant to the NLO
calculation for each contribution. They complete the NLO matching
condition for each Wilson coefficient at the electroweak scale $\mu_W$. 
In addition, we show in the table where we ignore the third
generation contribution. As we will see below, the effect of dropping
the NLO third-generation contribution is actually negligible.

%%%%%%%%%%%%%%%%%%%%%%%%%%%%%%%%%%%%%%%%%
\subsubsection{Higgs exchange}
%%%%%%%%%%%%%%%%%%%%%%%%%%%%%%%%%%%%%%%%

The Higgs exchange processes are induced by the effective coupling of
the wino DM with the Higgs boson. They only give the scalar-type
interactions as we show in Table~\ref{table:NLOcalc}.

In the case that the wino DM is close to the electroweak eigenstate,
the coupling is generated at one-loop level:
\begin{equation}
 {\cal L}_{{\chi}{\chi}h}= -\frac{1}{2}
c_{\rm H}(w)~
\bar{{\chi}}^0{\chi}^0 h^0~,
\end{equation}
where $w\equiv m_W^2/M^2$ with $m_W$ and $M$ being the masses of
$W$ boson and wino, respectively, and $c_{\rm H}(w)=
\frac{g_2^3}{(4\pi)^2}g_{\rm H}(w)$.  Here $g_{\rm H}(x)$ is a mass
function presented in Ref.~\cite{Hisano:2010fy}.\footnote{
  The mass functions used in text are collected in
  Appendix~\ref{app:massfunctions}. } By using the effective coupling
we readily obtain the LO matching condition for the scalar-type quark
operators as
\begin{equation}
 C^q_{\rm S}(\mu_{W}^{})\vert_{\rm LO}
  =\frac{\alpha_2^2}{4m_Wm^2_h}g_{\rm H}(w)~.
\end{equation}
Here $m_h$ is the mass of the Higgs boson and $\alpha_2\equiv
g_2^2/(4\pi)$. To evaluate the NLO matching condition, one needs to
evaluate the QCD corrections in the full and effective theories at
two- and one-loop levels, respectively. These corrections turn out to
be equivalent, and thus the matching condition does not differ from
the above equation, {\it i.e.},
\begin{equation}
 C^q_{\rm S}(\mu_{W}^{})=\frac{\alpha_2^2}{4m_Wm^2_h}g_{\rm H}(w)~,
\end{equation}
to the NLO in perturbation theory.

For the scalar-type gluon operator, the one-loop long-distance
contribution by the scalar-type quark operators is subtracted from the
two-loop contribution in the full theory so that only the top-quark
contribution is included in $C^G_{\rm S}$. Then, we have
\cite{Shifman:1978zn}
\begin{equation}
 C^G_{\rm S}(\mu_W^{})\vert_{\rm LO}
=-\frac{\alpha_2^2}{48 m_Wm_h^2}g_{\rm H}(w)~.
\end{equation}
At the NLO, the above expression is modified to \cite{Inami:1982xt,
Djouadi:2000ck} 
\begin{equation}
 C^G_{\rm S}(\mu_{W}^{})%\vert_{\rm NLO}
=-\frac{\alpha_2^2}{48 m_Wm_h^2}g_{\rm H}(w)
\biggl[
1+\frac{11}{4\pi}\alpha_s(\mu_{W}^{})
\biggr]
~.
\end{equation}
Notice that it contains no logarithmic terms like those containing a
factor of $\ln(m_t/\mu_{W}^{})$. This is
because $\frac{\alpha_s}{\pi} G^a_{\mu\nu}G^{a\mu\nu}$ is
renormalization-group invariant up to this order in perturbation theory.

%%%%%%%%%%%%%%%%%%%%%%%%%%%%%%%%%%%%%%%%%%%%%%%%%%%%%%%%%%
\subsubsection{Box type}
\label{sec:box}
%%%%%%%%%%%%%%%%%%%%%%%%%%%%%%%%%%%%%%%%%%%%%%%%%%%%%

Let us move on to the contribution of the box diagrams. They induce
both scalar-type and twist-2 operators. To compute the effective
operators, we first consider the OPEs of the correlation function of
the charged currents:
\begin{equation}
 \Pi_{\mu\nu}^W(q)\equiv i\int d^4xe^{iq\cdot
  x}~T\bigl\{J_\mu^W(x)J_\nu^W(0)^\dagger \bigr\} ~,
\end{equation}
where
\begin{equation}
 J^W_\mu \equiv \sum_{i=1,2,3}
\frac{g_2}{\sqrt{2}}
\overline{u}_i\gamma_\mu P_L d_i~,
\end{equation}
with $P_L\equiv (1-\gamma_5)/2$.
We evaluate the Wilson coefficients of the scalar and twist-2 operators
in the OPEs up to the NLO in $\alpha_s/\pi$.

We first consider the scalar part. It is convenient to decompose the
correlator into the transverse and the longitudinal parts as
\begin{equation}
 \Pi_{\mu\nu}^W(q)|_{\text{scalar}} 
= \biggl(-g_{\mu\nu}+\frac{q_\mu q_\nu}{q^2}\biggr)
\Pi^W_T(q^2)+\frac{q_\mu q_\nu}{q^2}\Pi^W_L(q^2)~,
\end{equation}
where
\begin{equation}
 \Pi^W_T(q^2)=\sum_{q}c^{q}_{W,S}(q^2;\mu_W)m_q\bar{q}q
+c^{G}_{W,S}(q^2;\mu_W)\frac{\alpha_s}{\pi} G^a_{\mu\nu}G^{a\mu\nu}~.
\end{equation}
Here we give only the transverse part since the longitudinal one does
not contribute to $C^q_{\rm S}$ and $C^G_{\rm S}$ \cite{Hisano:2010ct,
  Hisano:2011cs}.  As for the contribution to the scalar-type quark
operators of the first two generations, there is no ${\cal
  O}(\alpha_s^0)$ term since the charged current $J^W_\mu$ is pure
chiral (we take small quark mass limit for $q=u,d,s,c,b$). Thus, only
the one-loop diagrams are relevant in this case. It readily follows
from the results given in Ref.~\cite{Broadhurst:1994qj}, in which the
correlation functions for vector and axial currents are evaluated with
the OPEs, that
\begin{equation}
c^q_{W,S}(q^2;\mu_W)= -\frac{\alpha_s(\mu_{W}^{})}
{4\pi}\frac{g_2^2}{q^2}~,
\label{eq:NLOboxquarkscalar}
\end{equation}
for $q=u,d,s,c$.
On the other hand, the tree-level contribution of the bottom quark to the
scalar-type operator does not vanish because of the large top-quark
mass. We have
\begin{equation}
c^b_{W,S}(q^2;\mu_W)=
\frac{g_2^2m_t^2}{8(q^2_{}-m_t^2)^2}~.
%-\frac{\alpha_2\alpha_s(\mu_{W}^{})}{q^2}~.
\label{eq:cb_ws}
\end{equation}
Here, as mentioned above, we neglect the NLO contribution and take its
effects into account as a theoretical uncertainty.  The gluon
contribution of the first two generations is also obtained
straightforwardly from Ref.~\cite{Broadhurst:1994qj}. The contribution
of the third generation quarks, however, is not evaluated reliably by
means of the method used in Ref.~\cite{Broadhurst:1994qj} due to the
large mass of top quark.  Here again, we neglect the NLO effects and
consider them as a theoretical uncertainty. As a result, we obtain
\begin{equation}
 c^{G}_{W,S}(q^2;\mu_W)=
\frac{g_2^2}{48q^2}\biggl[2\times
\biggl(1+\frac{7}{6}\frac{\alpha_s(\mu_{W}^{})}{\pi}\biggr)
+
\biggl(
\frac{q^2}{q^2-m_t^2}
\biggr)
\biggr]
~,
\end{equation}
where the first and second terms in bracket correspond to the
contribution of the first two generations and the third generation,
respectively.

Next, we consider the twist-2 part. For the contribution of $q=u,d,s,c$
to the quark twist-2 operators, the relevant parts
are written as
\begin{align}
\Pi^W_{\mu\nu}(q)|_{(1,2)}^Q= \sum_{q=u,d,s,c}\frac{g_2^2}{2}
\biggl[-
\biggl(
\frac{g_{\mu\rho}g_{\nu\sigma}q^2-g_{\mu\rho}q_\nu q_\sigma 
- q_\mu q_\rho g_{\nu\sigma}
+g_{\mu\nu}q_\rho q_\sigma}{(q^2)^2}
\biggr)
c^q_{W,2}& 
\nonumber \\
+\biggl(g_{\mu\nu}-\frac{q_\mu q_\nu}{q^2}\biggr)
\frac{q_\rho q_\sigma}{(q^2)^2} 
c^q_{W,L}&
\biggr]{\cal O}^{q\rho\sigma}~.
\end{align}
The Wilson coefficients $c^q_{W,2}$ and $c^q_{W,L}$ are
evaluated in Refs.~\cite{Bardeen:1978yd} as follows: 
\begin{align}
 c_{W,2}^q(\mu_{W}^{})&=1+\frac{\alpha_s(\mu_{W}^{})}{4\pi}
\biggl[-\frac{1}{2}\biggl(\frac{64}{9}
\biggr)\ln\biggl(\frac{-q^2}{\mu_{W}^2}
\biggr)+\frac{4}{9}\biggr]~,\nonumber\\
 c_{W,L}^q(\mu_{W}^{})&=\frac{\alpha_s(\mu_{W}^{})}{4\pi}
\biggl[\frac{16}{9}\biggr]~.
\label{eq:twist2lightquark}
\end{align}
For the third generation contribution, on the other hand, we take into
account top mass in the LO part and neglect the NLO part as mentioned
above. As a result, we have
\begin{equation}
 \Pi^W_{\mu\nu}(q)|_{(3)}^Q= -\frac{g_2^2}{2}
\frac{1}{(q^2-m_t^2)^2}
\bigl[
(q^2-m_t^2)g_{\mu\rho}g_{\nu\sigma}
-g_{\mu\rho}q_\nu q_\sigma 
- q_\mu q_\rho g_{\nu\sigma}
+g_{\mu\nu}q_\rho q_\sigma
\bigr] c^b_{W,3}
{\cal O}^{b\rho\sigma}_{}~,
\end{equation}
with 
\begin{equation}
 c_{W,3}^b=1+\frac{\alpha_s(\mu_{W}^{})}{4\pi}
\biggl[-\frac{1}{2}\biggl(\frac{64}{9}
\biggr)\ln\biggl(\frac{-q^2}{\mu_{W}^2}
\biggr)\biggr]~.
\end{equation}
Note that we have included the logarithmic part though it is induced
at the NLO; otherwise, the Wilson coefficient shows wrong dependence
on the factorization scale $\mu_W$.  Finally let us derive the gluon
twist-2 operator.  It is always induced at ${\cal
  O}(\alpha_s/\pi)$. For the contribution of massless quarks, we use
the results given in Refs.~\cite{Bardeen:1978yd}. The result is
\begin{align}
\Pi^W_{\mu\nu}(q)|_{(1,2)}^G= \frac{g_2^2}{2}
\biggl[-
\biggl(
\frac{g_{\mu\rho}g_{\nu\sigma}q^2-g_{\mu\rho}q_\nu q_\sigma 
- q_\mu q_\rho g_{\nu\sigma}
+g_{\mu\nu}q_\rho q_\sigma}{(q^2)^2}
\biggr)
c^G_{W,2}& 
\nonumber \\
+\biggl(g_{\mu\nu}-\frac{q_\mu q_\nu}{q^2}\biggr)
\frac{q_\rho q_\sigma}{(q^2)^2} 
c^G_{W,L}&
\biggr]{\cal O}^{G\rho\sigma}~,
\end{align}
where
\begin{align}
  c_{W,2}^G(\mu_{W}^{})&=4\times \frac{\alpha_s(\mu_{W}^{})}{4\pi}
\biggl[-\frac{1}{2}\biggl(\frac{4}{3}
\biggr)\ln\biggl(\frac{-q^2}{\mu_{W}^2}
\biggr)+\frac{1}{2}\biggr]~,\nonumber\\
 c_{W,L}^G(\mu_{W}^{})&=4\times\frac{\alpha_s(\mu_{W}^{})}{4\pi}
\biggl[-\frac{2}{3}\biggr]~,
\end{align}
with a factor of four counting the number of the first two generation
quarks. As before, we neglect the NLO contribution of the third
generation quarks but keep its logarithmic part in order to guarantee
the appropriate scale dependence. This reads
\begin{align}
 \Pi^W_{\mu\nu}(q)|_{(3)}^G&= -\frac{g_2^2}{2}
\frac{1}{(q^2-m_t^2)^2}
\bigl[
(q^2-m_t^2)g_{\mu\rho}g_{\nu\sigma}
-g_{\mu\rho}q_\nu q_\sigma 
- q_\mu q_\rho g_{\nu\sigma}
+g_{\mu\nu}q_\rho q_\sigma
\bigr]\nonumber\\
&\times c_{W,3}^G
{\cal O}^{G\rho\sigma}_{}~,
\end{align}
with
\begin{equation}
 c_{W,3}^G = 
 \frac{\alpha_s(\mu_{W}^{})}{4\pi}
\biggl[-\frac{1}{2}\biggl(\frac{4}{3}
\biggr)\ln\biggl(\frac{-q^2}{\mu_{W}^2}
\biggr)\biggr] ~.
\end{equation}
Then, the sum of the above contributions gives the total twist-2
contribution: 
\begin{equation}
  \Pi^W_{\mu\nu}(q)|_{\text{twist2}} =  \Pi^W_{\mu\nu}(q)|_{(1,2)}^Q+
   \Pi^W_{\mu\nu}(q)|_{(3)}^Q
+ \Pi^W_{\mu\nu}(q)|_{(1,2)}^G+
   \Pi^W_{\mu\nu}(q)|_{(3)}^G  ~.
\end{equation}

Our remaining task is to obtain the Wilson coefficients of the effective
operators in Eq.~\eqref{eq:opdef} by computing another loop with the
electroweak current correlator $\Pi^W_{\mu\nu}(q)$. For
the scalar-type operators, we have
\begin{align}
C_{\rm S}^q(\mu_W) &=\frac{\alpha_2^2}{m_W^3}\frac{\alpha_s(\mu_W)}{4\pi}
[-12 g_{\rm B1}(w)]~,~~~~~~({\rm for}~~q=u,d,s,c)~,
%\label{eq:CqSBoxNLO}
\nonumber\\
C_{\rm S}^b(\mu_W) &=\frac{\alpha_2^2}{m_W^3}
\left[(-3)g_{\rm btm}(w,\tau)
%+ \frac{\alpha_s(\mu_W)}{4\pi} \{-12 g_{\rm B1}(x)\}
\right],
%\label{eq:CbSBoxNLO}
\nonumber\\
C_{\rm S}^G(\mu_W) &=\frac{\alpha_2^2}{4m_W^3}\left[
\left(2+\frac{7}{3}\frac{\alpha_s(\mu_W)}{\pi}\right)g_{\rm B1}(w)
+g_{\rm top}(w,\tau)
\right],
\label{eq:CgSBoxNLO}
\end{align}
where $\tau\equiv m_t^2/M^2$. 
The mass function $g_{\rm
  B1}(x)$ is given in Ref.~\cite{Hisano:2010fy}, and $g_{\rm
  top}(x,y)$ and $g_{\rm btm}(x,y)$ are equivalent to $g_{\rm
  B3}^{(1)}(x,y)$ and $g_{\rm B3}^{(2)}(x,y)$ in
Ref.~\cite{Hisano:2011cs}, respectively. These functions are also
presented in Appendix~\ref{app:massfunctions}.

For the twist-2 type operators, on the other hand, we have
\begin{align}
C^q_{\text{T}_i}(\mu_W)=&  \frac{\alpha_2^2}{m_W^3}
\left[g_{{\rm T}_i}(w,0)
+\frac{\alpha_s(\mu_W)}{4\pi}
\left( - \frac{32}{9}g_{{\rm T}_i}^{\rm log}(w,0;\mu_W)
+\frac{9}{4}g_{{\rm T}_i}(w,0)
+\frac{16}{9}h_{{\rm T}_i}(w)
\right)
\right]~,
\nonumber \\ 
C^b_{\text{T}_i}(\mu_W)=& \frac{\alpha_2^2}{m_W^3}\biggl[
g_{{\rm T}_i}(w,\tau)
+\frac{\alpha_s(\mu_W)}{4\pi}
\left(
- \frac{32}{9}g_{{\rm T}_i}^{\rm log}(w,\tau;\mu_W)
\right)
\biggr]~,\nonumber \\
C^G_{{\rm T}_i}(\mu_W) =&
\frac{\alpha_2^2}{m_W^3} \frac{\alpha_s(\mu_W)}{4\pi}\times \nonumber \\
&
\left[4\times
\biggl(
-\frac{2}{3}g_{{\rm T}_i}^{\rm log}(w,0;\mu_W)
+\frac{1}{2}g_{{\rm T}_i}(w,0)
- \frac{2}{3}h_{{\rm T}_i}(w)
\biggr)- \frac{2}{3}g_{{\rm T}_i}^{\rm log}(w,\tau;\mu_W)
\right]~,
\label{eq:wilsontw2ew}
\end{align}
where the functions $g_{{\rm T}_i}(x,y)$, $h_{{\rm T}_i}(x)$ and
$g_{{\rm T}_i}^{\rm log}(x, y;\mu_W)$ are given in
Appendix~\ref{app:massfunctions}. $g_{\text{T}_i}(x,0)$ agrees with
$g_{{\rm T}_i}(x)$ in, \textit{e.g.}, Ref.~\cite{Hisano:2010fy}.  The
terms proportional to $g^{\rm log}_{{\rm T}_i}(x,y)$ come from the
logarithmic terms in the OPEs of the correlation function of the
charged currents, while the terms with $g_{{\rm T}_i}(x,y)$ and
$h_{{\rm T}_i}(x)$ are from the non-logarithmic terms in
$c^{q/b/G}_{W,2}$ and $c^{q/G}_{W,L}$, respectively. The NLO
contribution to the gluon twist-2 operator is also given in
Ref.~\cite{Hill:2011be}.  Here we note that to obtain the proper
dependence of the above coefficients on the scale $\mu_W$, we need to
include all of the NLO corrections. Otherwise, the mismatch in the
scale dependence between the matching conditions and the RGEs causes
large uncertainties.

To that end, it is important to
appropriately perform the order counting with respect to
$\alpha_s/\pi$. Especially, the two-loop contribution to $C^G_{{\rm
T}_i}$ should be regarded as the NLO in $\alpha_s/\pi$,\footnote{One may
easily check that the logarithmic parts in the NLO contribution to the
twist-2 operators reproduce the one-loop RGEs presented in
Sec.~\ref{sec:RGE}. This justifies the order counting discussed here.}
not the LO, which is contrary to the case of the gluon scalar operator
$C^G_{\text{S}}$; in this case, the two-loop contribution is the LO in
$\alpha_s/\pi$. Our convention for the definition of the gluon operators
clarifies this order counting.

%%%%%%%%%%%%%%%%%%%%% Figure %%%%%%%%%%%%%%%%%%%%%%%%%%%%%%%%%%
\begin{figure}[t]
\begin{center}
 \includegraphics[clip, width = 0.5 \textwidth]{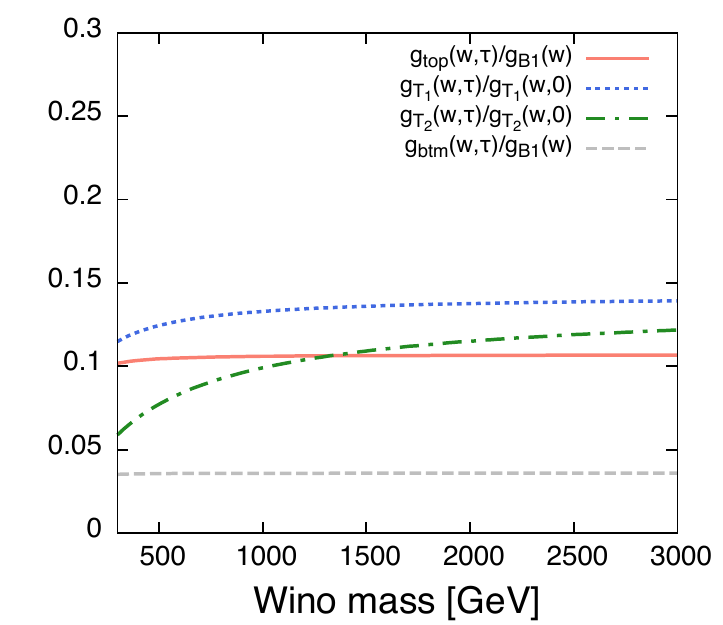}
\caption{Comparison of $g_{\text{top}}(w,\tau)$, $g_{\rm T_1}(w,
  \tau)$, $g_{\rm T_2}(w, \tau)$, and $g_{\text{btm}}(w, \tau)$ with
  $g_{\text{B1}}(w)$, $g_{\rm T_1}(w, 0)$, $g_{\rm T_2}(w, 0)$, and
  $g_{\text{B1}}(w)$ in red solid, blue dotted, green dash-dotted, and
  gray dashed lines, respectively, to show smallness of third
  generation contributions.  }
\label{fig:compare}
\end{center}
\end{figure}
%%%%%%%%%%%%%%%%%%%%%%%%%%%%%%%%%%%%%%%%%%%%%%%%%%%%%%%%%%%%%%%%

As we have already commented several times, we neglect the NLO
contribution of the third generation quarks. Indeed, we expect that
its significance is quite small, and thus we safely regard it as a
theoretical uncertainty.  In Fig.~\ref{fig:compare} we compare the
mass functions corresponding to the LO third generation contributions
with those of the LO massless quark contributions, which corresponds
to $g_{\text{top}}(w,\tau)/g_{\text{B1}}(w)$, $g_{\rm
  T_1}(w,\tau)/g_{\rm T_1}(w, 0)$, and $g_{\rm T_2}(w,\tau)/g_{\rm
  T_2}(w, 0)$.  $g_{\text{btm}}(w,\tau)/g_{\text{B1}}(w)$ is also
shown as its contribution to $C^G_{\text{S}}$ via integration of the
bottom quark is given by $-C^b_{\text{S}}/12$.  It is found that the
LO third generation contributions are smaller than those of the first
and second generations by almost an order of magnitude. Hence, we
expect that the NLO contributions of the third generation are also
considerably small compared with those of the other two
generations. This allows us to ignore the third-generation NLO
contribution, and treat it as a theoretical uncertainty.

%%%%%%%%%%%%%%%%%%%%%%%%%%%%%%%%%%%%%%%%%%%%%%%%%%%%%%%%%%%%%%%%
\subsection{Renormalization group equations and matching conditions}
\label{sec:RGE}
%%%%%%%%%%%%%%%%%%%%%%%%%%%%%%%%%%%%%%%%%%%%%%%%%%%%%%%%%%%%%%%

The effective operators are scale dependent and their scale evolution
is described by the RGEs. During the RG evolution, heavy quarks are
integrated out around their mass scale. Thus we need to match the
theories above and below the threshold.  Here we summarize the RGEs
and the matching conditions.

To begin with, we write down beta-function of $\alpha_s$ and
anomalous dimension of quark mass operator:
\begin{align}
  % \mu \frac{d \alpha_s}{d \mu}
 \beta(\alpha_s)
  &=(2b_1)\frac{\alpha_s^{2}}{4\pi}+(2b_2)\frac{\alpha_s^3}{(4\pi)^2} ~,
\label{eq:beta_s}
  \\
  \gamma_m&=-6C_F \frac{\alpha_s^{}}{4\pi}  ~,
  \label{eq:gamma_m}
\end{align} 
with 
%\begin{equation}
 $b_1=-\frac{11}3N_c+\frac23 N_f~,~
  b_2=-\frac{34}{3}N_c^2 +\frac{10}{3}N_cN_f+2C_F N_f~$.
%\label{b1b2}
%\end{equation}
($N_c=3$ is the number of colors, $N_f$ denotes the number of quark
  flavors in an effective theory and $C_F$ is the quadratic Casimir
  invariant defined by $C_F\equiv \frac{N_c^2-1}{2N_c}$.)  Here for
  the $\overline{\rm MS}$ quark masses, we use the one-loop anomalous
  dimension since their effects first appear at the NLO level as we
  will see below soon.

Now we give the RGEs for the Wilson coefficients of the above
operators. First, we consider the RGEs for the scalar-type
operators. To that end, notice that the quark mass operator is RG
invariant in a mass-independent renormalization scheme like the
$\overline{\rm MS}$ scheme, {\it i.e.},
\begin{equation}
\mu \frac{d}{d\mu}m_q\overline{q}q = 0~.
\end{equation}
To evaluate the evolution of the gluon scalar operator, we use the
trace anomaly formula \eqref{eq:traceanomaly}.  Differentiating
Eq.~\eqref{eq:traceanomaly}, we then obtain the differential equation
for the gluonic scalar operator $\frac{\alpha_s}{\pi}
G^a_{\mu\nu}G^{a\mu\nu}_{}$. As a result, we have\footnote{In fact, we
  implicitly assume that the operators are to be evaluated between the
  on-shell states. As discussed in Refs.~\cite{Inami:1982xt,
    Vecchi:2013iza}, during the RG flow, the scalar operators mix with
  other (gauge-variant) operators whose on-shell matrix elements
  vanish. }
\begin{equation}
 \mu \frac{d}{d\mu}(C^q_{\rm S}, C^G_{\rm S})=(C^q_{\rm S}, C^G_{\rm
  S})~\Gamma_{\rm S}^{}~, 
\end{equation}
where $\Gamma_{\rm S}$ is a $(N_f+1)\times (N_f+1)$ matrix given by
\begin{equation}
 \Gamma_{\rm S}=
\begin{pmatrix}
 0&\cdots&0&0\\
 \vdots& \ddots&\vdots&\vdots \\
0&\cdots& 0& 0\\
-4\alpha_s^2\frac{d \gamma_m}{d \alpha_s} & 
\cdots &
-4\alpha_s^2\frac{d \gamma_m}{d \alpha_s} & 
\alpha_s^2 \frac{d}{d \alpha_s}
\bigl(\frac{\beta(\alpha_s)}{\alpha_s^2}\bigr)
\end{pmatrix}
~,
\end{equation}
The solutions of the RGEs are given as follows:
\begin{align}
C^q_{\rm S}(\mu)&= C^q_{\rm S}(\mu_0)-4C^G_{\rm S}(\mu_0)
\frac{\alpha^2_s(\mu_0)}{\beta(\alpha_s(\mu_0))}
(\gamma_m(\mu)-\gamma_m(\mu_0)) ~,\label{eq:cqsrge} \\
C^G_{\rm S}(\mu)&=\frac{\beta(\alpha_s(\mu))}{\alpha_s^2(\mu)}
\frac{\alpha_s^2(\mu_0)}
 {\beta (\alpha_s(\mu_0))}C^G_{\rm S}(\mu_0)
~.
\end{align}
Eq.~\eqref{eq:cqsrge} shows that the anomalous dimension at ${\cal
  O}(\alpha_s)$, {\it i.e.} Eq.~\eqref{eq:gamma_m}, is enough for the
NLO calculation.

Next, we consider the RGEs for the twist-2 operators. The two-loop
anomalous dimension matrix of the operators is evaluated as
\cite{Floratos:1978ny, GonzalezArroyo:1979he}
\begin{equation}
 \mu \frac{d}{d\mu}(C^q_{\text{T}_i}, C^G_{\text{T}_i})=
(C^q_{\text{T}_i}, C^G_{\text{T}_i})~ \Gamma_{\rm T}~,
\end{equation}
with $\Gamma_{\rm T}$ a $(N_f+1)\times (N_f+1)$ matrix:
\begin{equation}
 \Gamma_{\rm T}=
\begin{pmatrix}
 \gamma_{qq}&0&\cdots&0&\gamma_{qg}\\
 0 &\gamma_{qq}&&\vdots&\vdots\\
\vdots&&\ddots&0&\vdots\\
0&\cdots&0&\gamma_{qq}&\gamma_{qg}\\
\gamma_{gq}&\cdots&\cdots&\gamma_{gq}&\gamma_{gg}
\end{pmatrix}
~,
\end{equation}
where
\begin{align}
 \gamma_{qq}&=\frac{16}{3}C_F\cdot\frac{\alpha_s}{4\pi}
+\biggl(-\frac{208}{27}C_FN_f-\frac{224}{27}C_F^2+\frac{752}{27}C_FN_c\biggr) 
\biggl(\frac{\alpha_s}{4\pi}\biggr)^2~,\nonumber \\
 \gamma_{qg}&=\frac{4}{3}\cdot\frac{\alpha_s}{4\pi}
+\biggl(\frac{148}{27}C_F+\frac{70}{27}N_c\biggr) 
\biggl(\frac{\alpha_s}{4\pi}\biggr)^2~,\nonumber \\
 \gamma_{gq}&=\frac{16}{3}C_F\cdot\frac{\alpha_s}{4\pi}
+\biggl(-\frac{208}{27}C_FN_f-\frac{224}{27}C_F^2+\frac{752}{27}C_FN_c\biggr) 
\biggl(\frac{\alpha_s}{4\pi}\biggr)^2~,\nonumber \\
 \gamma_{gg}&=\frac{4}{3}N_f\cdot\frac{\alpha_s}{4\pi}
+\biggl(\frac{148}{27}C_FN_f+\frac{70}{27}N_cN_f\biggr) 
\biggl(\frac{\alpha_s}{4\pi}\biggr)^2~.
\end{align}

%%%%%%%%%%%%%%%%%%%%%%%%%%%%%%%%%%%%%%%
%\subsection{Quark  threshold}
%\label{sec:quarkthr}
%%%%%%%%%%%%%%%%%%%%%%%%%%%%%%%%%%%%%%

%When we across a quark threshold during the RGE evolution,
Finally we give the threshold corrections at the scale where heavy
quarks are integrated out.  For example, in the vicinity of the
bottom-quark threshold $\mu_b\simeq m_b$, we match the strong gauge
coupling constant and the Wilson coefficients as
\begin{equation}
 \frac{1}{\alpha_s(\mu_b)|_{N_f=4}}
= \frac{1}{\alpha_s(\mu_b)|_{N_f=5}}
+\frac{1}{3\pi}\ln\biggl(\frac{\mu_b}{m_b}\biggr) ~,
\end{equation}
and
\begin{align}
 C^q_{\rm S}(\mu_b)|_{N_f=4}&= C^q_{\rm S}(\mu_b)|_{N_f=5}~,\nonumber \\
[\alpha_s C^G_{\rm S}](\mu_b)|_{N_f=4}&=
-\frac{\alpha_s(\mu_b)}{12}
\left[
1+\frac{\alpha_s(\mu_b)}{4\pi}
\left(11+\frac{2}{3}\ln\frac{m_b^2}{\mu_b^2}\right)
\right]
C^b_{\rm S}(\mu_b)|_{N_f=5}
\nonumber \\ &
+
\left[
1+\frac{\alpha_s(\mu_b)}{4\pi}
\frac{2}{3}
\ln\frac{m_b^2}{\mu_b^2}
\right]
[\alpha_s C^G_{\rm S}](\mu_b)|_{N_f=5}~,\nonumber \\
% C^G_{\rm S}(\mu_b)|_{N_f=4}&=
% -\frac{1}{12\pi} \biggl[1+\frac{11}{4\pi}\alpha_s(\mu_b)\biggr]
% C^b_{\rm S}(\mu_b)|_{N_f=5}+ 
%  C^G_{\rm S}(\mu_b)|_{N_f=5}~,\nonumber \\
 C^q_{\text{T}_i}(\mu_b)|_{N_f=4}&= C^q_{\text{T}_i}
(\mu_b)|_{N_f=5}~,\nonumber \\
 C^G_{\text{T}_i}(\mu_b)|_{N_f=4}&= 
\left[
1+\frac{\alpha_s(\mu_b)}{4\pi}
\frac{2}{3}
\ln\frac{m_b^2}{\mu_b^2}
\right]
C^G_{\text{T}_i}(\mu_b)|_{N_f=5}
+\frac{\alpha_s(\mu_b)}{4\pi}\frac{2}{3}\ln \frac{m_b^2}{\mu_b^2}
C^b_{\text{T}_i}(\mu_b)|_{N_f=5}
~,
\label{eq:quarkthmatch}
\end{align}
with $q=u,d,s,c$ for the first and third equations.\footnote{The
matching condition for $C^G_{\text{T}_i}$ here differs from that given
in Ref.~\cite{Hill:2011be}.} In the following section, we estimate
the uncertainties coming from the neglect of the higher order
perturbation by varying the matching scale $\mu_b$ around the
$\mu_b\simeq m_b$.  We repeat a similar procedure for the charm-quark
threshold around $\mu_c\simeq m_c$.

Here we note that besides the above threshold corrections, the higher
dimension operators suppressed by a power of the threshold quark mass
are also generated in general. For instance, if the scalar-type quark
operator is integrated out at a quark threshold $m_Q$, then we will
obtain the following dimension-nine operators at one-loop level
\cite{Cho:1994yu, Vecchi:2013iza}:
\begin{equation}
 -\frac{\alpha_s(m_Q)}{60\pi m_Q^2} (D^\nu G^a_{\nu\mu}) (D^\rho
  G^a_{\rho \mu})\bar{\chi}^0\chi^0
-\frac{g_s\alpha_s(m_Q)}{720\pi m_Q^2}
f_{abc} G^a_{\mu\nu} G^{b\mu\rho} G^c_{\nu\rho} \bar{\chi}^0\chi^0~,
\label{eq:dimsix}
\end{equation}
where $f_{abc}$ is the SU(3) structure constant. In particular, those
generated at the charm-quark threshold give the largest effects. By
using the naive dimensional analysis, we see that their contribution
to the nucleon matrix element may give a correction by a factor of
$\Lambda^2_{\text{QCD}}/m_c^2 = {\cal O}(0.1)$, which could be
additionally suppressed by the prefactors of these operators. Since we
do not know precise values of the nucleon matrix elements of the
operators in Eq.~\eqref{eq:dimsix}, we should also consider their
effects as an uncertainty.

%%%%%%%%%%%%%%%%%%%%%%%%%%%%%%%%%%%%%%
%\section{Error estimate}
\section{Results}
\label{sec:error}
%%%%%%%%%%%%%%%%%%%%%%%%%%%%%%%%%%%

%%%%%%%%%%%%%%%%%%%%%%%%%%%%%%%%%%%%%%%%%%%%%%%%
\begin{table}[t]
\caption{Input parameters.}
\label{table:inputparameters}
 \begin{center}
\begin{tabular}{l|l}
\hline
\hline
Strong coupling constant $\alpha_s(m_Z)$ \cite{Agashe:2014kda} &
 $0.1185\pm 0.0006$ \\
Higgs pole mass $m_h$ \cite{Aad:2014aba, Khachatryan:2014ira} & $125.03\pm
     0.27$~GeV \\ 
Top-quark pole mass $m_t$ \cite{ATLAS:2014wva} & $173.34\pm0.76$~GeV \\
\hline
\hline
\end{tabular}
\end{center}
\end{table}
%%%%%%%%%%%%%%%%%%%%%%%%%%%%%%%%%%%%%%%%%%%%%%%%%%%%%%%%%

Now we compute the wino-nucleon scattering cross section and evaluate
the theoretical uncertainties. We first separately consider the scalar
and twist-2 contributions to the wino-nucleon effective coupling in
Sec.~\ref{sec:scalarfp} and \ref{sec:twist2fp},
respectively. Then, we show the result for the scattering cross
section in the following subsection. In
Table~\ref{table:inputparameters}, we summarize the input parameters
we use in our computation. For the mass of top quark, we use the pole
mass as an input parameter, and convert it to the
$\overline{\text{MS}}$ mass using the one-loop relation:
\begin{equation}
 m_t =  \overline{m}_t(\overline{m}_t) \biggl[1+\frac{4\alpha_s
  (\overline{m}_t) }{3\pi}\biggr] ~, 
\end{equation} 
where $\overline{m}_t$ denotes the $\overline{\text{MS}}$ top mass. In
what follows, we only use the $\overline{\text{MS}}$ mass so we drop the
bar for brevity.

%%%%%%%%%%%%%%%%%%%%%%%%%%%%%%%%%
\subsection{Scalar part}
\label{sec:scalarfp}
%%%%%%%%%%%%%%%%%%%%%%%%%%%%%%%

%%%%%%%%%%%%%% FIGURE %%%%%%%%%%%%%%%%%%%%%%%%%%%%%%%%%%%%
\begin{figure}[t!]
\begin{center}
\subfigure[Perturbation]
 {\includegraphics[clip, width = 0.48 \textwidth]{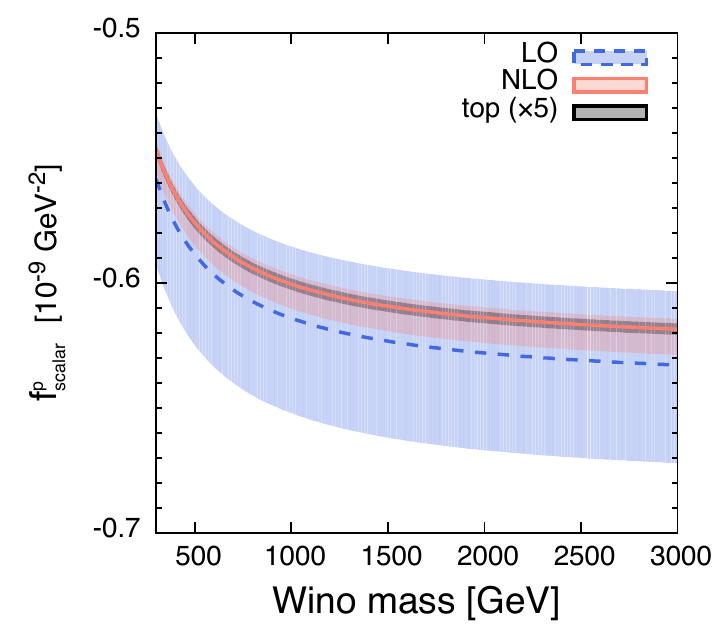}
 \label{fig:scalarpert}}
\hspace{-0.04\textwidth}
\subfigure[Input]
 {\includegraphics[clip, width = 0.48 \textwidth]{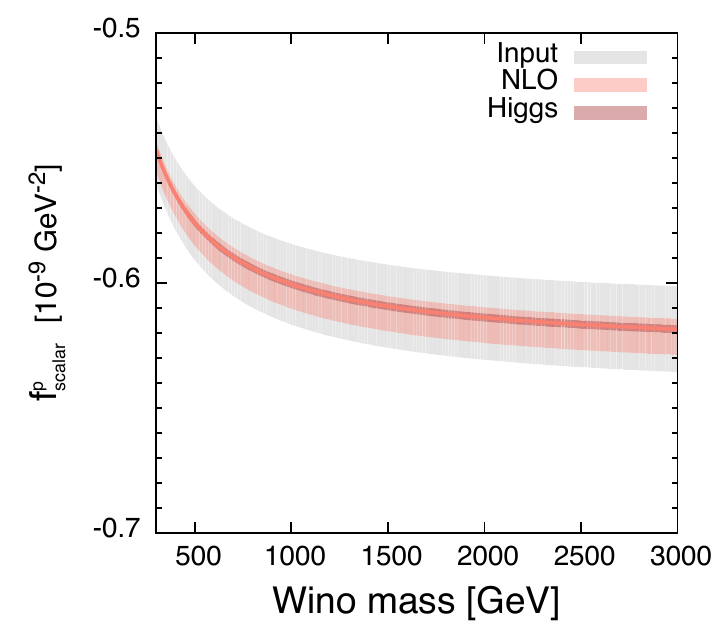}
 \label{fig:scalarinperr}}\\
\subfigure[OPE]
 {\includegraphics[clip, width = 0.48 \textwidth]{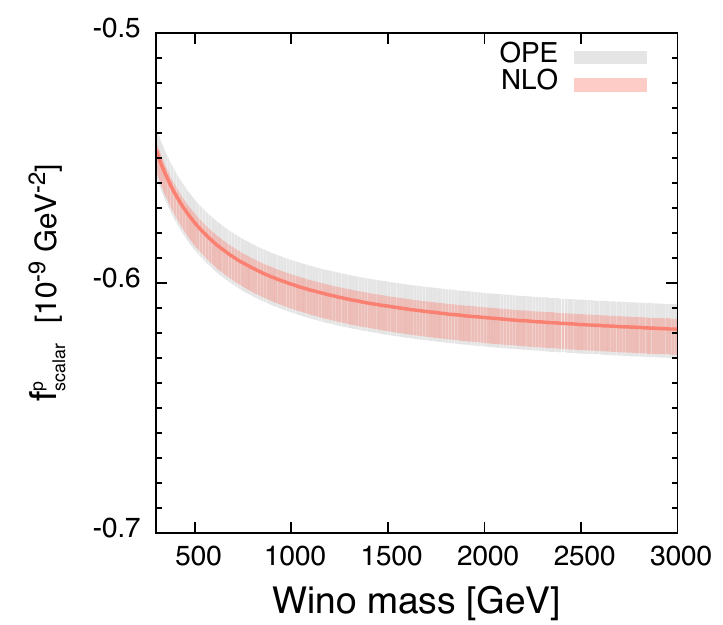}
 \label{fig:scalaropeerr}}
\caption{Contribution of scalar-type operators to wino-proton coupling
  $f^p_{\text{scalar}}$.  (a) LO (blue dashed) and NLO (red solid)
  results with corresponding bands showing uncertainty due to
  perturbative calculation.  Gray band indicates uncertainty coming
  from lack of NLO contribution of third generation, multiplied by a
  factor of five. (b) Errors from input parameters (gray), the Higgs
  mass (dark red), compared with NLO error (pink).  (c) Uncertainty
  from truncating higher dimension operators at each quark threshold
  (gray band), compared with NLO perturbative QCD uncertainty (pink
  band).}
\label{fig:scalar}
\end{center}
\end{figure}
%%%%%%%%%%%%%%%%%%%%%%%%%%%%%%%%%%%%%%%%%%%%%%%%%%%%%%%%%%

The spin-independent effective coupling of the wino with nucleon is
defined by
\begin{equation}
 {\cal L}^{(N)}_{\text{SI}} =f^N \bar{\chi}^0 \chi^0 \overline{N} N ~.
\label{eq:sinucleoncoupling}
\end{equation}
The contribution of the scalar operators to the coupling is given by
\begin{equation}
 f^N_{\text{scalar}} =\sum_{q=u,d,s}C^q_{\text{S}}(\mu_{\text{had}})
\langle N|m_q\bar{q}q|N\rangle +C^G_{\text{S}}(\mu_{\text{had}})
\langle N|\frac{\alpha_s}{\pi} G^a_{\mu\nu}G^{a\mu\nu}|N\rangle ~,
\end{equation}
where we take the hadron scale $\mu_{\text{had}}=1$~GeV with $N_f =3$
active quarks.  Fig.~\ref{fig:scalar} shows $f^p_{\text{scalar}}$ with
various types of errors.

In Fig.~\ref{fig:scalar} (a) $f^p_{\text{scalar}}$ at the LO (blue
dashed) and NLO (red solid) with corresponding bands showing the
theoretical error due to the perturbative calculation are shown.  In
the plot the uncertainty coming from lack of the NLO contribution of
the third generation, which is multiplied by a factor of five just for
the purpose of presentation, is also shown (gray band). For the
evaluation of the error from the ignorance of higher order
contribution in perturbation, we vary each matching scale by a factor
of two; {\it i.e.}, $m_c/2\leq \mu_c \leq 2m_c$, $m_b/2\leq \mu_b \leq
2m_b$, $m_Z/2\leq \mu_W \leq 2m_Z$. The prescription is, however, less
effective for the scalar-type operators since these operators are
almost scale-invariant. For this reason, when evaluating the error
resulting from the quark threshold matching for the NLO (LO)
calculation, we use the NNLO (NLO) matching conditions to artificially
generate the logarithmic dependence of the Wilson coefficients on the
scale by using the mismatch between the matching conditions and
RGEs. The NLO matching conditions are given in
Eq.~\eqref{eq:quarkthmatch}, while the NNLO ones are found in
Ref.~\cite{Chetyrkin:1997un}.  In addition, for the LO contribution,
we evaluate the uncertainty caused by the electroweak-scale matching
by merely multiplying the LO contribution by a factor of
$\alpha_s/\pi$. Since the scalar-type operators are scale-invariant at
the LO, it is impossible to estimate the LO uncertainty from the
electroweak-scale matching by varying the scale $\mu_W$. At the NLO,
on the other hand, we are to estimate the uncertainty with the scale
variation since the NLO RGEs yield the scale dependence of the scalar
operators.

The error from the LO perturbative calculation is more than 5\%, which
reduces to a few \% level with the NLO calculation. The upper errors
smaller than the lower errors in the LO and NLO perturbative
calculations in the Fig.~~\ref{fig:scalar} (a). This comes from
difference between $\alpha_s(m_Q/2)$ and $\alpha_s(2m_Q)$ for $Q=b,c$.
On the other hand, as for the uncertainty due to the lack of the
third-generation NLO contribution, we estimate its effect by
multiplying the LO contribution by a factor of $\alpha_s/\pi$. From
the figure, we find that the ignorance of the third-generation NLO
contribution only gives a negligible effect on the resultant
value. The effect is much smaller than the uncertainty due to the
perturbative calculation.

%Fig.~\ref{fig:scalarinperr}
Fig.~\ref{fig:scalar} (b)
shows comparison of the uncertainty in the
NLO perturbative QCD calculation (pink) with that from the errors in
the input parameters we have used in the calculation (gray).  Among
them, the uncertainty coming from the Higgs mass error is especially
shown in the dark red band. We see that thanks to the NLO calculation
the perturbative error now becomes smaller than the error from the
input parameters, though they are still of the same order of the
magnitude.

Finally we plot the theoretical uncertainty which could arise due to
the higher dimension operators induced at each quark threshold in
%Fig.~\ref{fig:scalaropeerr}.
Fig.~\ref{fig:scalar} (c). To evaluate the effects of the higher
dimension operators, we vary the scalar gluon contribution induced at
the charm-quark threshold by $2$\%, which is expected from the naive
dimensional analysis as discussed in Sec.~\ref{sec:RGE}.\footnote{
Since the first (second) operator in Eq.~\eqref{eq:dimsix} receives
additional suppression by a factor of five (sixty) compared with the
contribution of the scalar gluon operator, $-\alpha_s/(12\pi)
GG\bar{\chi}^0\chi^0$, we estimate the significance of the former
contribution as $\sim 2$\% of that of the scalar-type gluon operator,
while the latter contribution is negligible.
} Since the higher dimension operators generated at the bottom-quark
threshold are suppressed by the bottom quark mass, their effects are
negligible. As seen from the figure, this uncertainty may be as large
as the NLO perturbative QCD error. To reduce the uncertainty, one of
the most efficient ways is to use the nucleon matrix elements computed
above the charm-quark threshold, say, at the scale of 2~GeV. In this
case, we need to evaluate the charm-quark content in nucleon,
$f_{T_c}^{(N)}=\langle N|m_c\bar{c}c| N\rangle /m_N$, as
well. Currently, the QCD lattice simulations are not able to compute
it accurately \cite{Dinter:2012tt}. If this quantity is evaluated with
good precision in the future, then the uncertainty due to the
higher dimension operators will be significantly reduced.  We expect
that the perturbative QCD error will also decrease, since we do not
need the charm-quark threshold matching procedure any more. Thus, we
strongly encourage the development in this field.

%%%%%%%%%%%%%%%%%%%%%%%%%%%%%%%
\subsection{Twist-2 part}
\label{sec:twist2fp}
%%%%%%%%%%%%%%%%%%%%%%%%%%%%%%

%%%%%%%%%%%%%% FIGURE %%%%%%%%%%%%%%%%%%%%%%%%%%%%%%%%%%%%
\begin{figure}[t!]
\begin{center}
\subfigure[$g_{{\text{T}_1}}^{\rm log}(w,0;\mu_W)$]
 {\includegraphics[clip, width = 0.48 \textwidth]{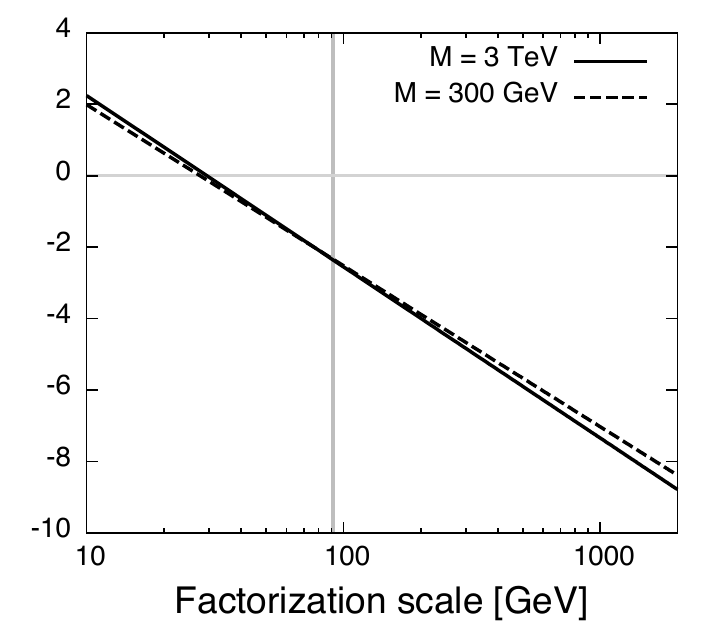}
 \label{fig:gt1log}}
\hspace{-0.04\textwidth}
\subfigure[$g_{{\text{T}_2}}^{\rm log}(w,0;\mu_W)$]
 {\includegraphics[clip, width = 0.48 \textwidth]{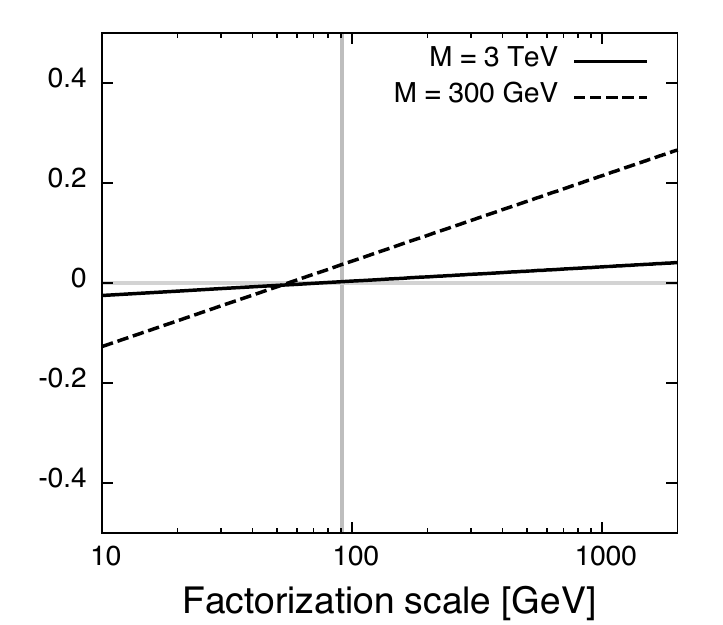}
 \label{fig:gt2log}}
\caption{$g_{{\text{T}_i}}^{\rm log}(w,0;\mu_W)$ $(i=1,2)$ as function
  of factorization scale $\mu_W$. $M=3$~TeV (solid) and $300$~GeV
  (dashed).  Vertical gray line shows $\mu_W = m_Z$.  }
\label{fig:twist2log}
\end{center}
\end{figure}
%%%%%%%%%%%%%%%%%%%%%%%%%%%%%%%%%%%%%%%%%%%%%%%%%%%%%%%%%%

Contrary to the scalar-type operators, the twist-2 operators have the
scale dependence at the leading order in $\alpha_s$. Therefore, it is
necessary to determine the appropriate scale for the matching of the
full theory onto the effective theory in order not to suffer from
large logarithmic factors. To that end, we require that the
logarithmic dependent parts $g^{\text{log}}_{\text{T}_i}$ in the
Wilson coefficients presented in Eq.~\eqref{eq:wilsontw2ew} should not
be large, say, within ${\cal O}(1)$. Since the terms proportional to
$g^{\text{log}}_{\text{T}_i}$ come from the logarithmic terms in the
OPEs of the correlation function of the charged currents, this
condition guarantees the validity of the perturbative QCD
expansion. In Fig.~\ref{fig:twist2log}, we show $g_{{\text{T}_i}}^{\rm
  log}(w,0;\mu_W)$ $(i=1,2)$ as function of the factorization scale
$\mu_W$. Here $M=3$~TeV (solid) and $300$~GeV (dashed).
The vertical gray line shows $\mu_W = m_Z$. It turns out that the size
of these functions is within ${\cal O}(1)$ if one takes the scale
$\mu_W$ to be around the electroweak scale.  This consequence rarely
depends on the DM mass. The absolute values for these functions are
minimum at a scale of ${\cal O}(10)$~GeV, which is much smaller than
the DM mass.  This observation reflects the fact that the typical
scale of the loop momentum flowing in the loop diagrams in
Fig.~\ref{fig:DiagramTot} is around the electroweak scale, as pointed
out in Ref.~\cite{Hisano:2004pv}. In the following calculation, we
take $\mu_W = m_Z$, which assures that $g^{\text{log}}_{\text{T}_i}$
is within ${\cal O}(1)$ and thus the perturbative expansion is
justified.

To calculate the contribution of the twist-2 operators, we also need
to choose the scale at which the nucleon matrix elements of the
twist-2 operators are evaluated. As mentioned above, contrary to the
case of the scalar-type operators, the twist-2 matrix elements are
obtained at various scales. Since the result does not depend on the
choice of the scale within the uncertainty of the calculation, it is
desirable to choose the scale so that the error in calculation is
reduced. Thus, we take it to be the same as the factorization scale,
\textit{i.e.}, $\mu = m_Z$. This choice allows us to decrease the
error which would arise from the process where the operators are
evolved down to the low-energy region; for instance, if one evaluates
the matrix elements at a scale $\mu < m_b$, the result suffers from
the uncertainty resulting from the bottom-quark mass threshold. See
Ref.~\cite{Hisano:2015bma} for further discussion.

%%%%%%%%%%%%%% FIGURE %%%%%%%%%%%%%%%%%%%%%%%%%%%%%%%%%%%%
\begin{figure}[t!]
\begin{center}
\subfigure[Perturbation]
 {\includegraphics[clip, width = 0.48 \textwidth]{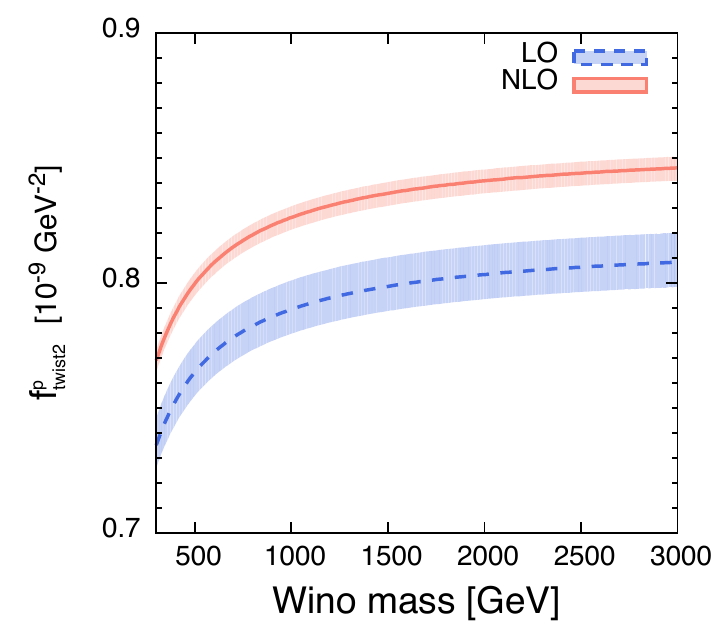}
 \label{fig:twist2pert}}
\hspace{-0.04\textwidth}
\subfigure[Input]
 {\includegraphics[clip, width = 0.48 \textwidth]{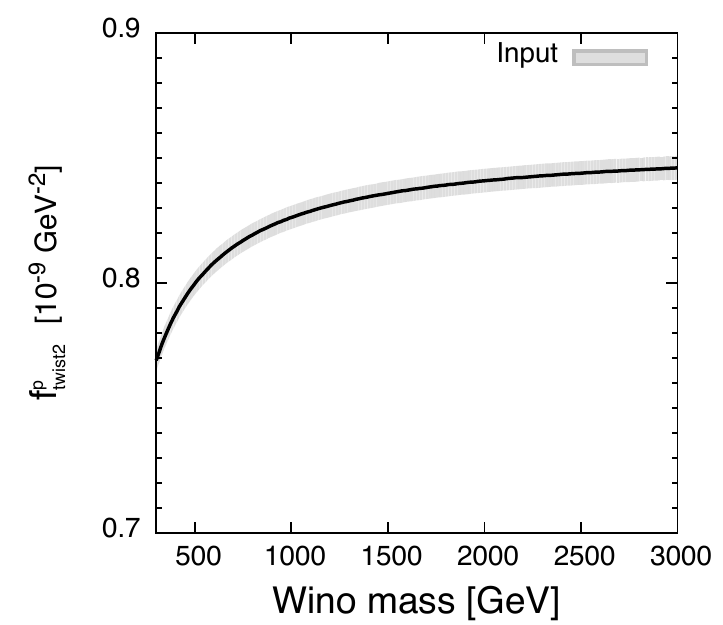}
 \label{fig:twinperr}}
\caption{Contribution of twist-2 operators to wino-proton coupling
  $f^p_{\text{twist2}}$.  (a) LO (blue dashed) and NLO (red solid)
  results with corresponding bands showing uncertainty due to
  perturbative calculation.  (b) Uncertainty resulting from input
  error. }
\label{fig:twist2}
\end{center}
\end{figure}
%%%%%%%%%%%%%%%%%%%%%%%%%%%%%%%%%%%%%%%%%%%%%%%%%%%%%%%%%%

Now we evaluate the contribution of the twist-2 operators to the SI
effective coupling in Eq.~\eqref{eq:sinucleoncoupling}, which is given by
\begin{align}
 \frac{f^N_{\text{twist2}}}{m_N} &= \frac{3}{4}\sum_{q}
\sum_{i=1,2}C^q_{\text{T}_i}(m_Z)
[q^{(N)}(2;m_Z)+\bar{q}^{(N)}(2;m_Z)]\nonumber\\
& -\frac{3}{4}
\sum_{i=1,2}C^G_{\text{T}_i}(m_Z)g^{(N)}(2; m_Z) ~,
\end{align}
where $q$ runs over the active quarks ($q=u,d,s,c,b$ for our choice of
the scale $\mu = m_Z$). 

In Fig.~\ref{fig:twist2}, we show $f^p_{\text{twist2}}$ as function of
the wino mass. We compare the LO and NLO results in the left panel,
shown in the blue dashed and red solid lines, respectively, with the
corresponding bands representing the uncertainties. The uncertainties
are evaluated by varying the scale $\mu_W$ between $m_Z/2$ and
$2m_Z$. Besides, it is found that to drop the NLO contribution of the
third generation quarks causes only the negligible effects, so we do
not show the error due to the contribution. The ${\cal O}(1)$\% error
in the LO computation now reduces to $\sim 0.5$\% when going to the
NLO level, though the central value shifts more than expected, {\it
  i.e.}  about $5$\% change. This is due to a large NLO term in
$C^q_{{\rm T}_i}$ of Eq.\,\eqref{eq:wilsontw2ew}. In the large DM mass
limit, the contributions of quarks and gluon at the NLO are $0.90$ and
$-0.047$ in $10^{-9}\,{\rm GeV}^{-2}$ unit, respectively, while the
quark contribution at the LO is $0.82$ in $10^{-9}\,{\rm GeV}^{-2}$
unit.\footnote{To be concrete, in $C^q_{{\rm T}_i}$ the NLO term
  summed over $i=1,\,2$ gives $(\alpha_2^2\alpha_s/4\pi m_W^3) \times (41\pi/12)$ in the
  large DM mass limit. Here logarithmic term $g^{\rm log}_{{\rm T}_i}$
  is neglected for simplicity. See also
  Eqs.\,\eqref{eq:Lim_g_H}--\eqref{eq:Lim_f_A^num} for the mass functions in the
large DM mass limit.} In the right panel
of Fig.~\ref{fig:twist2}, we also illustrate the uncertainty resulting
from the input error, which turns out to be as large as the NLO
uncertainty. The uncertainty mainly comes from those of the PDFs,
which we estimate following the method given in
Ref.~\cite{Owens:2012bv} with the $\chi^2$ tolerance $T$ taken to be
$T=10$. After all, in the case of the twist-2 contribution, both the
NLO and input uncertainties are less than 1\%, and thus well
controlled compared to the scalar contribution.

%%%%%%%%%%%%%%%%%%%%%%%%%%%%%%%%%%%%%%%
\subsection{Scattering cross section}
%%%%%%%%%%%%%%%%%%%%%%%%%%%%%%%%%%%%%%%

%%%%%%%%%%%%%%%%%%%%% Figure %%%%%%%%%%%%%%%%%%%%%%%%%%%%%%%%%%
\begin{figure}[t]
\begin{center}
 \includegraphics[clip, width = 0.6 \textwidth]{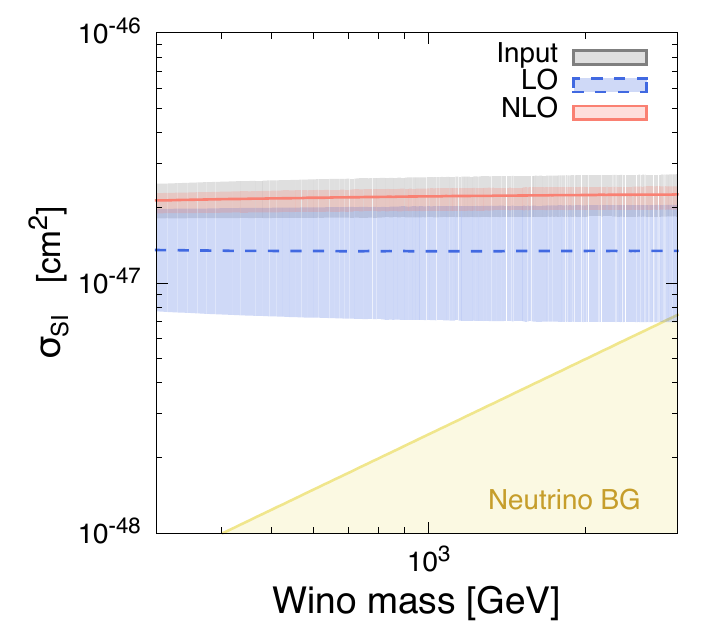}
\caption{Wino-proton SI scattering cross section.  Blue dashed and
  red solid lines represent LO and NLO results, respectively, with
  corresponding bands show perturbative uncertainties. Gray band shows
  uncertainty resulting from the input error. Yellow shaded area
  corresponds to the region in which neutrino background overcomes DM
  signal \cite{Billard:2013qya}.  }
\label{fig:sicross}
\end{center}
\end{figure}
%%%%%%%%%%%%%%%%%%%%%%%%%%%%%%%%%%%%%%%%%%%%%%%%%%%%%%%%%%%%%%%%

Finally, we evaluate the wino-nucleon SI scattering cross section,
which is given by
\begin{equation}
 \sigma^N_{\text{SI}} =\frac{4}{\pi}\biggl(\frac{M m_N}{M+m_N}\biggr)^2
|f^N_{\text{scalar}}+f^N_{\text{twist2}}|^2 ~.
\end{equation}
We plot $\sigma^p_{\text{SI}}$  as function of the wino mass in
Fig.~\ref{fig:sicross}.
Additionally we indicate the parameter region where the neutrino
background dominates the the DM-nucleon scattering
\cite{Billard:2013qya} and then it becomes hard to detect the DM signal
in the DM direct detection experiments (yellow shaded).
Here we estimate each error by varying the scalar and twist-2
contributions within their uncertainties evaluated above.
The result shows that the large uncertainty in the LO computation is
significantly reduced once the NLO QCD corrections are included, which
is now smaller than that from the input error. In the large DM mass
limit, the SI scattering cross section converges to a constant value,
\begin{equation}
 \sigma_{\text{SI}}^p = 2.3~{}^{+0.2}_{-0.3} ~{}^{+0.5}_{-0.4}\times 10^{-47}~
  \text{cm}^2 ~,
\end{equation}
where the first and second terms represent the perturbative and input
uncertainties, respectively. As seen from Fig.~\ref{fig:sicross},
$\sigma_{\text{SI}}^p$ has little dependence on the DM mass; its
variation is actually within the uncertainties of the calculation, for
the wino mass larger than $270$~GeV. Both the scalar and twist-2
contributions depend on the DM mass when the mass is smaller than
$\sim 1$~TeV as shown in Figs.~\ref{fig:scalar} and
\ref{fig:twist2}. However, the dependence in the cross section is
accidentally canceled. The NLO result is found to be larger than the
LO result by almost 70\%. After all, the resultant scattering cross
section is well above that of the neutrino background
\cite{Billard:2013qya}, and therefore the future direct detection
experiments are promising to test the wino DM scenario.

%%%%%%%%%%%%%%%%%%%%%%%%%%%%%%%%%%%%%%%%%
\section{Electroweakly-interacting DM}
\label{sec:ewimp}
%%%%%%%%%%%%%%%%%%%%%%%%%%%%%%%%%%%%%%%

Although we have focused on the wino DM in this paper, a similar
formalism may be constructed for a more general class of the DM
candidates; {\it i.e.}, an SU(2)$_L$ multiplet with hypercharge $Y$
that contains a neutral component for DM, and their thermal relic may
explain the observed DM density with ${\cal O}(1)$~TeV masses. For
previous works on such DM candidates, see Refs.~\cite{Cirelli:2005uq,
  Essig:2007az, Hambye:2009pw, Hisano:2014kua, Nagata:2014wma,
  Nagata:2014aoa, Abe:2014gua, Boucenna:2015haa}. Some theories beyond
the Standard Model actually predict this kind of DM. For example, the
higgsino and wino in the SUSY models are representative of the
SU(2)$_L$ multiplet DM. Moreover, such a particle may show up in grand
unified theories \cite{Kadastik:2009dj, Frigerio:2009wf,
  Mambrini:2013iaa}, whose stability is explained by a remnant
discrete symmetry of extra U(1) symmetries in the theories
\cite{Krauss:1988zc, Ibanez:1991hv, Martin:1992mq, DeMontigny:1993gy,
  Mambrini:2015vna}.

Before concluding our discussion, we give the results of the NLO
calculation for this class of DM candidates. If the DM particle is a
fermion, its interactions with quarks and gluon are completely
determined by the electroweak gauge interactions,\footnote{In the case
  of the scalar DM, on the other hand, there always exist quartic
  couplings to the Higgs boson, and the couplings also induce the
  interactions of the DM with quarks and gluon.} so we consider the
fermionic DM candidates in the following discussion.  If $Y\neq 0$,
the DM is a Dirac fermion, while a Majorana fermion if $Y=0$. Pure
Dirac fermion DM is, however, severely constrained by the direct
detection experiments already, since the vector interactions via the
$Z$ boson exchange yield too large scattering cross section with
nucleon. The constraint may be evaded if there are some new physics
effects that give rise to the mass difference between the neutral
components to split them into two Majorana fermions. If the mass
difference is larger than ${\cal O}(100)$~keV, the scatterings with
nucleon are not induced by the tree-level $Z$ boson exchange. In what
follows, we assume the presence of the mass difference and regard the
lighter neutral component $\chi^0$ as a DM candidate. The mass
difference is assumed to be small enough to be neglected in the
following calculation.  In this case, the interactions including the
neutral components are given by
\begin{align}
 \mathcal{L}_{\text{int}}
&= \frac{g_2}{4}\sqrt{n^2-(2Y-1)^2}~
\overline{\chi^+}\Slash{W}^+\chi^0
+\frac{g_2}{4}\sqrt{n^2-(2Y+1)^2}~\overline{\chi^0}\Slash{W}^+\chi^-
+{\rm h.c.}
\nonumber \\
&+{ig_ZY}\overline{\chi^0}\Slash{Z}\eta^0~. 
\end{align}
Here $n$ is the number of the components in the DM SU(2)$_L$
multiplet, $g_Z\equiv \sqrt{g_Y^2+g_2^2}$ with $g_Y$ the U(1)$_Y$
gauge coupling constant, and $\eta^0$ and $Z_\mu$ for the heavier
neutral component and the $Z$ boson, respectively.

The LO calculation of the scattering cross section with a nucleon for
this type of DM candidates is given in Ref.~\cite{Hisano:2011cs}. As
in the case of the wino DM, we find that there is a significant
cancellation among the contributions to the scattering
amplitude. Therefore, the NLO corrections are of importance to
evaluate the scattering cross section precisely.  We compute the NLO
scattering cross section in a similar manner to above discussion. The
only difference is the electroweak matching conditions, which we
summarize in Appendix~\ref{sec:ewimpmatching}. Below the electroweak
scale, the procedure is completely the same as before.

%%%%%%%%%%%%%%%%%%%%% Figure %%%%%%%%%%%%%%%%%%%%%%%%%%%%%%%%%%
\begin{figure}[t]
\begin{center}
 \includegraphics[clip, width = 0.6 \textwidth]{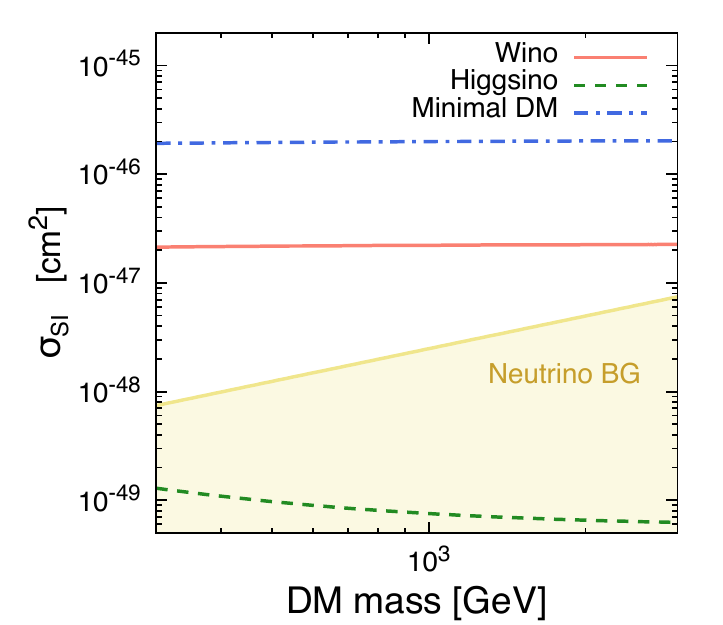}
\caption{SI scattering cross sections of the SU(2)$_L$ multiplet DM
  candidates. Red solid, green dashed, and blue dash-dotted lines
  correspond to the $(n, Y)=(3,0)$, $(2,1/2)$, and $(5, 0)$ cases,
  respectively. Yellow shaded area indicates the region in which
  neutrino background overcomes the DM signal \cite{Billard:2013qya}.
}
\label{fig:ewsicross}
\end{center}
\end{figure}
%%%%%%%%%%%%%%%%%%%%%%%%%%%%%%%%%%%%%%%%%%%%%%%%%%%%%%%%%%%%%%%%

In Fig.~\ref{fig:ewsicross} we plot the SI scattering cross sections
for several SU(2)$_L$ multiplet DM candidates. Here the red solid,
green dashed, and blue dash-dotted lines represent the $(n, Y)=(3,0)$,
$(2,1/2)$, and $(5, 0)$ cases, respectively. The triplet case
corresponds to the wino DM, while the doublet one is regarded as the
higgsino DM. The $(n, Y)=(5,0)$ fermion DM is the so-called minimal DM
\cite{Cirelli:2005uq}, for which the gauge symmetry guarantees its
stability. Again, the yellow shaded area indicates the region in which
neutrino background overcomes the DM signal \cite{Billard:2013qya}. We
find that all of the scattering cross sections are almost constant in
the mass region we are interested in, as already seen in the case of
wino DM. In the heavy DM mass limit, the DM-proton effective coupling
$f^p\equiv f^p_{\text{scalar}}+f^p_{\text{twist2}}$ at the NLO is
given by
\begin{equation}
 f^p =(n^2-4Y^2-1)f^p_W +Y^2 f^p_Z ~,
\end{equation}
with
\begin{align}
 f^p_W &= 2.9\times 10^{-11} ~\text{GeV}^{-2} ~, \nonumber \\
 f^p_Z &= -1.8 \times 10^{-10} ~\text{GeV}^{-2} ~,
\end{align}
from which one readily obtains the SI scattering cross section for a
generic SU(2)$_L$ DM candidate.  It is seen that the $(n, Y)=(3,0)$
and $(5,0)$ cases offer the SI scattering cross sections well above
the neutrino background, while that of the $(n, Y)=(2,1/2)$ case falls
far below the background. Compared to the previous results in
Ref.~\cite{Hisano:2011cs}, slightly larger SI scattering cross
sections are obtained for DM candidates with $Y=0$. As for the $(n,
Y)=(2,1/2)$ case, on the other hand, we obtain a smaller SI scattering
cross section.\footnote{Here we note that the LO $Z$ boson
  contribution is found to be different from that given in
  Ref.~\cite{Hisano:2011cs} by a factor of two.}

%%%%%%%%%%%%%%%%%%%%%%%%%%%%%%%%%%%%%%
\section{Conclusion and discussion}
\label{sec:conclusion}
%%%%%%%%%%%%%%%%%%%%%%%%%%%%%%%%%%%%%%%

In this paper we have completed the calculation of the wino-nucleon
scattering cross section up to the NLO in $\alpha_s/\pi$. It turns out
that the inclusion of the NLO corrections allows us to reduce the
theoretical uncertainty significantly, which is now ${\cal O}(10)$\%
level. The NLO scattering cross section is larger than the LO one by
about 70\%. The resultant cross section is well above the neutrino
background, and thus the DM direct detection experiment is a promising
tool for examining the wino DM scenario. In addition, we give the NLO
results for the cases with a generic SU(2)$_L$ multiplet DM, some of
which may also be probed in future experiments.

At present, the uncertainties from the input parameters, especially those
of the scalar matrix elements, dominate the theoretical error. If future
lattice simulations determine the charm-quark content in nucleon with
good accuracy, the uncertainties are to be reduced considerably.
We strongly anticipate the developments in the field.

%%%%%%%%%%%%%%%%%%%%%%%%%%%%%%%%%%%%
\section*{Acknowledgments}
%%%%%%%%%%%%%%%%%%%%%%%%%%%%%%%%%%%%

We would like to thank Hiroshi Ohki for useful discussions.  The work
is supported by Grant-in-Aid for Scientific research from the Ministry
of Education, Science, Sports, and Culture (MEXT), Japan, No. 24340047
and No. 23104011, (J.H.), Research Fellowships of the Japan Society
for the Promotion of Science for Young Scientists (N.N.), and also by
World Premier International Research Center Initiative (WPI
Initiative), MEXT, Japan (J.H.and N.N.). This work was supported
in part by the German Science Foundation (DFG) within the
Collaborative Research Center 676 Particles, ``Strings and the Early
Universe''.

%%%%%%%%%%%%%%%%%%%%%%%%%%%%%%%%%%%%%%%%%%%%%%
\section*{Appendix}
\appendix
%%%%%%%%%%%%%%%%%%%%%%%%%%%%%%%%%%%%%%%%%%%%%

%%%%%%%%%%%%%%%%%%%%%%%%%%%%%%%%%%%%%%%%%%%%
\section{Mass functions}
\setcounter{equation}{0} 
\label{app:massfunctions}
%%%%%%%%%%%%%%%%%%%%%%%%%%%%%%%%%%%%%%%%%%%

Here we list the mass functions used in text:
\begin{align}
 g_{\rm H}(x)&= 2\sqrt{x}(2-x\ln x)-\frac{2}{b_x}(2+2x-x^2)\tan^{-1}\biggl(
\frac{2b_x}{\sqrt{x}}\biggr)~,\\
 g_{\rm B1}(x)&=\frac{1}{24}\sqrt{x}(2-x\ln x)+\frac{1}{24b_x}
(4-2x+x^2){\rm tan}^{-1}\biggl(\frac{2b_x}{\sqrt{x}}\biggr)
~,\\
 g_{\rm btm}(x,y)&=-\frac{x^{\frac{3}{2}}y}{12(x-y)^2}
-\frac{x^{\frac{5}{2}}y^2}{24(x-y)^3}\ln\biggl(\frac{x}{y}\biggr)
\nonumber \\
&-\frac{xy(2y+6x+2xy-x^2y)}{24b_x(x-y)^3}\tan^{-1}
\biggl(\frac{2b_x}{\sqrt{x}}\biggr)
\nonumber \\
&+\frac{x^{\frac{3}{2}}y^{\frac{1}{2}}(2x+6y+2xy-xy^2)}{24b_y(x-y)^3}\tan^{-1}
\biggl(\frac{2b_y}{\sqrt{y}}\biggr)~,
\\
g_{\rm
 top}(x,y)&=\frac{x^{\frac{3}{2}}}{12(x-y)}-\frac{x^{\frac{5}{2}}(x-2y)}
 {24(x-y)^2}\ln x -\frac{x^{\frac{3}{2}}y^2}{24(x-y)^2}\ln y
\nonumber \\
&+\frac{x\{x^3+4y+4x(1+y)-2x^2(1+y)\}}{24b_x(x-y)^2}\tan^{-1}
\biggl(\frac{2b_x}{\sqrt{x}}\biggr) \nonumber \\
&-\frac{x^{\frac{3}{2}}y^{\frac{1}{2}}b_y(2+y)}{6(x-y)^2}\tan^{-1}
\biggl(\frac{2b_y}{\sqrt{y}}\biggr)~,
\\
g_{{\rm T_1}}(x,y) &=
\frac{x^{\frac{3}{2}}\{x(1-2x)+y(13+2x)-2y^2\}}{12(x-y)^2} 
\nonumber \\ 
&-
\frac{x^{\frac{3}{2}}\{x^3(2-x)+2xy(3-3x+x^2)+6y^2(2-x)\}}{12(x-y)^3}
\ln x
\nonumber \\  &+
\frac{x^{\frac{3}{2}}y\{2x(3-6y+y^2)+y(12+2y-y^2)\}}{12(x-y)^3}
\ln y
\nonumber \\
&+ 
\frac{x\{4x^2b_x^2(2+x^2)-2xy(6-7x+5x^2-x^3)-6y^2(2-4x+x^2)\}}{12b_x(x-y)^3}
\tan^{-1}\left(\frac{2b_x}{\sqrt{x}}\right)
\nonumber \\ &-
\frac{x^{\frac{3}{2}}y^{\frac{1}{2}}\{2x(3-y)(2+5y-y^2)-y(2-y)(14+2y-y^2)\}}{12b_y(x-y)^3} 
\tan^{-1}\left(\frac{2b_y}{\sqrt{y}}\right)
,
\\
 g_{{\rm T_2}}(x,y) &=
 \frac{x^{\frac{3}{2}}\{x(-1+2x)-(1+2x)y+2y^2\}}{4(x-y)^2}
\nonumber \\ &+
\frac{x^{\frac{5}{2}}\{(2-x)x^2+2y(1-3x+x^2)\}}{4(x-y)^3}
\ln x
\nonumber \\ &+
\frac{x^{\frac{3}{2}}y\{y^2(y-2)-2x(1-3y+y^2)\}}{4(x-y)^3}
\ln y
 \nonumber \\ &+
\frac{x^3\{x(2-4x+x^2)-2y(5-5x+x^2)\}}{4b_x(x-y)^3}
\tan^{-1}\left(\frac{2b_x}{\sqrt{x}}\right)
\nonumber \\ &+
\frac{x^{\frac{3}{2}}y^{\frac{3}{2}}(2x(5-5y+y^2)-y(2-4y+y^2))}{4b_y(x-y)^3}
\tan^{-1}\left(\frac{2b_y}{\sqrt{y}}\right)~,
 \\ 
 h_{\rm T_1}(x)&=-\frac{\sqrt{x}}{12}\{3-2x-x(3-x)\ln x\}
+\frac{1}{3}b_x~x(1-x)\tan^{-1}\biggl(
\frac{2b_x}{\sqrt{x}}
\biggr)~,\\
 h_{\rm T_2}(x)&=-\frac{\sqrt{x}}{12}
\{1+6x+x(4-3x)\ln x\}+
\frac{1}{12b_x}(4-2x+10x^2-3x^3)\tan^{-1}\biggl(
\frac{2b_x}{\sqrt{x}}
\biggr)
~,
\end{align}
where we have defined $b_x\equiv \sqrt{1-x/4}$. Note that 
\begin{align}
 g_{\rm T_1}(x,0)&=\frac{1}{12}\sqrt{x}\{1-2x-x(2-x)\ln x\}
+\frac{1}{3}b_x(2+x^2)\tan^{-1}\biggl(\frac{2b_x}{\sqrt{x}}\biggr)
~,\\
g_{\rm T_2}(x,0)&=-\frac{1}{4}\sqrt{x}\{1-2x-x(2-x)\ln x\}+
\frac{1}{4b_x}x(2-4x+x^2)\tan^{-1}\biggl(\frac{2b_x}{\sqrt{x}}\biggr)~,
\end{align}
are equal to $g_{{\rm T}_1}(x)$ and $g_{{\rm T_2}}(x)$ in
Ref.~\cite{Hisano:2010fy}, respectively. 
On the other hand, $g^{\rm log}_{\text{T}_i}(x,y;\mu_W)$ are
given by the following integrals:
\begin{equation}
 g^{\rm log}_{\text{T}_i}(x,y;\mu_W)=g^{\rm num}_{\text{T}_i}(x,y)
+\ln \biggl(x \frac{M^2}{\mu^2_W}\biggr)g_{\text{T}_i}(x,y)~,
\end{equation}
with
\begin{align}
 g^{\rm num}_{\rm T_1}(x,y)=&\frac{x^{\frac{3}{2}}}{24}
\int^{\infty}_{0}dt~
\frac{1}{(t+x)^2(t+y)^2}
\biggl[6y\bigl\{-4t-t^2+(2+t)\sqrt{t}\sqrt{4+t}\bigr\}
\nonumber \\
&+t\bigl\{-6t+4t^2+t^3+(2-t)(4+t)\sqrt{t}\sqrt{4+t}\bigr\}
\biggr]\ln\biggl(
\frac{t}{x}\biggr)
~, \nonumber \\[3pt]
 g^{\rm num}_{\rm T_2}(x,y)=&\frac{x^{\frac{3}{2}}}{8}
\int^{\infty}_{0}dt~
\frac{t^2\{-2-4t-t^2+(2+t)\sqrt{t}\sqrt{4+t}
\}}{(t+x)^2(t+y)^2}\ln\biggl(
\frac{t}{x}\biggr)~.
\end{align}
We compute these integrals numerically.

For the generic SU(2)$_L$ DM case, we further introduce the following
functions: 
\begin{align}
  f_V(x,y)&= f_V^{\text{anl}}(x,y)+f_V^{\text{num}}(x,y) ~, \nonumber \\
  f_A(x,y)&= f_A^{\text{anl}}(x,y)+f_A^{\text{num}}(x,y) ~,
\end{align}
where
\begin{align}
 f_V^{\text{anl}}(x,y)&=-\frac{\sqrt{x}(x^2-xy+12y^2)}{12(x-4y)^2} \nonumber \\
&+\frac{x^{\frac{3}{2}}(x^3-12x^2y+20xy^2 -48y^3)}{24(x-4y)^3}\ln x
+\frac{x^{\frac{3}{2}}y^2(7x-4y)}{6(x-4y)^3}\ln(4y) \nonumber \\
&+\frac{x^{\frac{3}{2}}y^{\frac{1}{2}}\{5x+28y+2y(7x-4y)(1-2y)\}}
{12(x-4y)^3\sqrt{1-y}}\tan^{-1}\biggl(\frac{\sqrt{1-y}}{\sqrt{y}}\biggr) 
\nonumber \\
&-\frac{4(x^3+44xy^2-48y^3)+x(x-2)(x^3-12x^2y+20xy^2-48y^3)}{24(x-4y)^3b_x}
\tan^{-1}\biggl(\frac{2b_x}{\sqrt{x}}\biggr) ~,\\
 f_A^{\text{anl}}(x,y)&=
\frac{\sqrt{x}(x-2y)}{4(x-4y)}
-\frac{x^{\frac{3}{2}}(x^2-8xy+8y^2)}{8(x-4y)^2}\ln x
-\frac{x^{\frac{3}{2}}y^2}{(x-4y)^2}\ln(4y) \nonumber \\
&+\frac{x^{\frac{3}{2}}\sqrt{y}(2y^2-y-1)}{(x-4y)^2\sqrt{1-y}}
\tan^{-1}\biggl(\frac{\sqrt{1-y}}{\sqrt{y}}\biggr) ~ \nonumber \\
&+\frac{4(x^2-2xy+8y^2)+x(x-2)(x^2-8xy+8y^2)}
{8(x-4y)^2b_x}
\tan^{-1}\biggl(\frac{2b_x}{\sqrt{x}}\biggr) ~,
\end{align}
while $f_V^{\text{num}}(x,y)$ and $f_A^{\text{num}}(x,y)$ are expressed
by the integral form as
\begin{align}
 f_V^{\text{num}}(x,y)&=-x^{\frac{3}{2}}y^2\int^\infty_{0}dt
\frac{(t+2y)\{(2-t)\sqrt{t+4}+t\sqrt{t}\}}{2t(t+x)^2(t+4y)^{\frac{5}{2}}}
\ln\biggl(\frac{\sqrt{t+4y}+\sqrt{t}}{\sqrt{t+4y}-\sqrt{t}}\biggr) ~, \\
 f_V^{\text{num}}(x,y)&=x^{\frac{3}{2}}y^2\int^\infty_{0}dt
\frac{(t+4y)\{(2-t)\sqrt{t+4}+t\sqrt{t}\}}{2t(t+x)^2(t+4y)^{\frac{5}{2}}}
\ln\biggl(\frac{\sqrt{t+4y}+\sqrt{t}}{\sqrt{t+4y}-\sqrt{t}}\biggr) ~.
\end{align}
Again, these integrals are evaluated numerically. The functions
$f_V^{\text{anl}}(x,y)$ and $f_A^{\text{anl}}(x,y)$ are given 
by functions in Ref.~\cite{Hisano:2011cs} as 
$f_V^{\text{anl}}(x,y)=G_{t1}(x,y)/4$ and $f_A^{\text{anl}}(x,y)=G_{t2}(x,y)/4$.

In the large DM mass limit, \textit{i.e.}, $x,y\to 0$ with the ratio
$y/x$ fixed, the above analytic functions are reduced to as follows:
\begin{align}
  g_{\text{H}}(x)&\to -2\pi ~,
  \label{eq:Lim_g_H}\\
 g_{\text{B1}}(x) &\to \frac{\pi}{12} ~, \\
 g_{\text{btm}}(x,y) &\to \frac{\pi}{24}\frac{r}
{(1+r)^3} ~, \\[3pt]
 g_{\text{top}}(x,y) &\to \frac{\pi}{12(1+r)^2} ~, \\[3pt]
 g_{\text{T}_1}(x,y) &\to \frac{\pi(2+3r)}{6(1+r)^3} ~, \\[3pt]
 g_{\text{T}_2} (x,y) &\to 0 ~, \\[3pt]
 h_{\text{T}_1} (x) &\to 0 ~, \\[3pt]
 h_{\text{T}_2} (x) &\to \frac{\pi}{6} ~, \\[3pt]
 g^{\rm num}_{{\rm T}_1}(x,y) &\to  -\frac{\pi\{
(1+r)^2(1-r)(2-3r)+(3-7r^2)r\ln r\}}{3(1-r^2)^3} ~, \\[3pt]
 g^{\rm num}_{{\rm T}_2}(x,y) &\to 0~, \\[3pt]
f_V^{\text{anl}}(x,y) &\to \frac{\pi}{24}\frac{(-2+5r+28r^3-88r^4 +96r^6)}
{(1-4r^2)^3}~, \\[3pt]
f_A^{\text{anl}}(x,y) &\to \frac{\pi}{4}\frac{(1-2r-2r^2+8r^4)}
{(1-4r^2)^2}~, %\\[3pt]
\end{align}
with $r\equiv \sqrt{y/x}$ and
\begin{align}
  f_V^{\text{num}}(z,\tau) &\to -0.189~, \\[3pt]
  f_A^{\text{num}}(z,\tau) &\to 0.364~.
\label{eq:Lim_f_A^num} 
\end{align}
Here we have set the values for the masses of $Z$ boson and top quark
in $z$ and $\tau$, respectively. 

%%%%%%%%%%%%%%%%%%%%%%%%%%%%%%%%%%%%%%%%%%%%%%%%%%%%%%%%%%%%%%%%%%%%%
\section{Results for the electroweak-interacting DM}
\label{sec:ewimpmatching}
%%%%%%%%%%%%%%%%%%%%%%%%%%%%%%%%%%%%%%%%%%%%%%%%%%%%%%%%%%%%%%%%%%%

In this Appendix, we summarize the electroweak matching conditions for
generic SU(2)$_L$ multiplet DM. 

%%%%%%%%%%%%%%%%%%%%%%%%%%%%%%%%%%%%%%%%%%%
\subsection{Current correlator}
%%%%%%%%%%%%%%%%%%%%%%%%%%%%%%%%%%%%%%%%%%%

To begin with, we consider the OPEs of the electroweak current
correlators as in Sec.~\ref{sec:box}. The correlation function
of the charged currents has been already discussed there. Here we
give the OPEs of the neutral current correlator, for it is necessary
to evaluate the $Z$ boson contribution. The correlation function of
the weak neutral current is defined by
\begin{equation}
 \Pi_{\mu\nu}^Z(q)\equiv i\int d^4x~ e^{iq\cdot
  x}T\{J_\mu^Z(x)J_\nu^Z(0)^\dagger \}~,
\end{equation}
where
\begin{equation}
 J_\mu^Z=\frac{g_Z}{2}\sum_{q}\overline{q} \gamma^\mu 
(g_V^q-g_A^q\gamma^5)q ~,
\end{equation}
with 
\begin{equation}
 g_V^q\equiv T^3_{q_L}-2\sin^2\theta_W Q_q ~,~~~~~~
 g_A^q\equiv T^3_{q_L} ~.
\end{equation}

Let us first evaluate the Wilson coefficients of the scalar operators.
For the scalar operators,  
the correlator is decomposed to the transverse and longitudinal parts as 
\begin{equation}
 \Pi_{\mu\nu}^Z(q)|_{\text{scalar}}
 = \biggl(-g_{\mu\nu}+\frac{q_\mu q_\nu}{q^2}\biggr)
\Pi^Z_T(q^2)+\frac{q_\mu q_\nu}{q^2}\Pi^Z_L(q^2)~.
\end{equation}
Again, only the transverse part is relevant to the calculation. The OPE
coefficients are defined by
\begin{equation}
 \Pi^Z_T(q^2)=\sum_{q}c^{q}_{Z,S}(q^2;\mu_W)m_q\bar{q}q
+c^{G}_{Z,S}(q^2;\mu_W)\frac{\alpha_s}{\pi} G^a_{\mu\nu}G^{a\mu\nu}~,
\end{equation}
are then evaluated as follows \cite{Broadhurst:1994qj}:
\begin{align}
 c^q_{Z,S}(q^2;\mu_W) &=\frac{g_Z^2}{2q^2}
\biggl[\{(g_V^q)^2 -(g_A^q)^2\} +
\frac{\alpha_s}{3\pi}\{(g_V^q)^2 -7(g_A^q)^2\}
\biggr] ~, \\[3pt]
 c^G_{Z,S}(q^2;\mu_W) &=\sum_{q}
\frac{g_Z^2}{48 q^2}\biggl(1+\frac{7\alpha_s}{6\pi}\biggr)
\{(g_V^q)^2 +(g_A^q)^2\} \nonumber \\
&+
\frac{g_Z^2\{(g_V^t)^2(-q^4+4m_t^2q^2-12m_t^4)
+3(g_A^t)^2(q^2-4m_t^2)(q^2-2m_t^2)\}}{48q^2(q^2-4m_t^2)^2} \nonumber \\
&+
\frac{g_Z^2m_t^4\{(g_V^t)^2(q^2-2m_t^2)-(g_A^t)^2(q^2-4m_t^2)\}
\sqrt{1-\frac{4m_t^2}{q^2}}~\ln\biggl(
\frac{\sqrt{1-\frac{4m_t^2}{q^2}}+1}{\sqrt{1-\frac{4m_t^2}{q^2}}-1}
\biggr)
}{4q^2
(q^2-4m_t^2)^3} ~,
\end{align}
with $q=u,d,s,c,b$. Here we drop the NLO contribution of top quark for
simplicity. This contribution is also readily obtained from the results in
Ref.~\cite{Broadhurst:1994qj}. The LO
terms in the above equations agree with the results given in
Ref.~\cite{Hisano:2011cs}.

Next, we consider the twist-2 operators. Their contribution to the
correlation function is written as \cite{Bardeen:1978yd}
\begin{align}
\Pi^Z_{\mu\nu}(q)|_{\text{twist2}} =g_Z^2\sum_{i=q,G}
\biggl[-
&\biggl(
\frac{g_{\mu\rho}g_{\nu\sigma}q^2-g_{\mu\rho}q_\nu q_\sigma 
- q_\mu q_\rho g_{\nu\sigma}
+g_{\mu\nu}q_\rho q_\sigma}{(q^2)^2}
\biggr)
c^i_{Z,2} 
\nonumber \\
&+\biggl(g_{\mu\nu}-\frac{q_\mu q_\nu}{q^2}\biggr)
\frac{q_\rho q_\sigma}{(q^2)^2} 
c^i_{Z,L}
\biggr]{\cal O}^{i\rho\sigma}~,
\end{align}
with the coefficients given by
\begin{align}
 c^q_{Z,2}(\mu_W) &=[(g_V^q)^2 +(g_A^q)^2]
\biggl\{ 1+\frac{\alpha_s(\mu_W)}{4\pi}
\biggl[-\frac{1}{2}\biggl(\frac{64}{9}\biggr)\ln \biggl(
\frac{-q^2}{\mu_W^2}\biggr)+\frac{4}{9}\biggr]\biggr\} ~, \nonumber \\[3pt]
 c^q_{Z,L}(\mu_W) &=[(g_V^q)^2 +(g_A^q)^2]
\biggl\{ \frac{\alpha_s(\mu_W)}{4\pi}
 \biggl[\frac{16}{9}\biggr]\biggr\}  ~, \nonumber \\[3pt]
 c^G_{Z,2}(\mu_W) &=\sum_{q=u,d,s,c,b}
[(g_V^q)^2 +(g_A^q)^2]
\biggl\{ \frac{\alpha_s(\mu_W)}{4\pi} 
\biggl[-\frac{1}{2}\biggl(\frac{4}{3} \biggr)\ln \biggl(
\frac{-q^2}{\mu_W^2}\biggr)+\frac{1}{2}\biggr]\biggr\}  ~, \nonumber \\[3pt]
 c^G_{Z,L}(\mu_W) &=\sum_{q=u,d,s,c,b}[(g_V^q)^2 +(g_A^q)^2]
\biggl\{ \frac{\alpha_s(\mu_W)}{4\pi} 
\biggl[-\frac{2}{3}\biggr]\biggr\}  ~.
\end{align}
Here again, we have neglected the top-quark contribution to the NLO gluon
coefficients.

%%%%%%%%%%%%%%%%%%%%%%%%%%%%%%%%%%%%%%%%%%
\subsection{Wilson coefficients}
%%%%%%%%%%%%%%%%%%%%%%%%%%%%%%%%%%%%%%%%%%

Now we calculate the electroweak-scale matching conditions. For the
scalar-type quark operators, we have 
\begin{align}
 C^q_{\text{S}}(\mu_W) &= \frac{\alpha_2^2}{4m_h^2}
\biggl[\frac{n^2-(4Y^2+1)}{8m_W}g_{\text{H}}(w)
+\frac{Y^2}{2m_Z\cos^4\theta_W}g_{\text{H}}(z) 
\biggr] \nonumber \\[3pt]
&+\frac{\alpha_2^2}{m_W^3}\frac{n^2-(4Y^2+1)}{8}\frac{\alpha_s(\mu_W)}{4\pi}
[-12g_{\text{B1}}(w)] \nonumber \\[3pt]
&+\frac{\alpha_2^2Y^2}{\cos^4\theta_W m_Z^3}
\biggl[\{(g_{V}^q)^2 -(g_{A}^q)^2\} +
\frac{\alpha_s}{3\pi}\{(g_V^q)^2 -7(g_A^q)^2\}
\biggr][3g_{\text{B1}}(z)] ~,
\end{align}
for $q=u,d,s,c$, and 
\begin{align}
 C^b_{\text{S}}(\mu_W) &= \frac{\alpha_2^2}{4m_h^2}
\biggl[\frac{n^2-(4Y^2+1)}{8m_W}g_{\text{H}}(w)
+\frac{Y^2}{2m_Z\cos^4\theta_W}g_{\text{H}}(z) 
\biggr] \nonumber \\[3pt]
&+\frac{\alpha_2^2}{m_W^3}\frac{n^2-(4Y^2+1)}{8}
[(-3)g_{\text{btm}}(w,\tau)] \nonumber \\[3pt]
&+\frac{\alpha_2^2Y^2}{\cos^4\theta_W m_Z^3}
\biggl[\{(g_V^q)^2 -(g_A^q)^2\} +
\frac{\alpha_s}{3\pi}\{(g_V^q)^2 -7(g_A^q)^2\}
\biggr][3g_{\text{B1}}(z)] ~,
\end{align}
where $\theta_W$ is the weak mixing angle and $z\equiv
m_Z^2/M^2$.\footnote{Note that $g_{\text{S}}(x)$ in
Ref.~\cite{Hisano:2011cs} is equal to $6g_{\text{B1}}(x)$. } 
The Wilson coefficient of the scalar-type gluon operator is, on the
other hand, computed as
\begin{align}
  C^G_{\text{S}}(\mu_W) &=- \frac{\alpha_2^2}{48m_h^2}
\biggl[
1+\frac{11}{4\pi}\alpha_s(\mu_{W}^{})
\biggr]
\biggl[\frac{n^2-(4Y^2+1)}{8m_W}g_{\text{H}}(w)
+\frac{Y^2}{2m_Z\cos^4\theta_W}g_{\text{H}}(z) 
\biggr] \nonumber \\[3pt]
&+\frac{\alpha_2^2}{4m_W^3}
\frac{n^2-(4Y^2+1)}{8}\left[
\left(2+\frac{7}{3}\frac{\alpha_s(\mu_W)}{\pi}\right)g_{\rm B1}(w)
+g_{\rm top}(w,\tau)
\right] \nonumber \\
& +\frac{\alpha_2^2Y^2}{8\cos^4\theta_W m_Z^3}
\biggl[
\sum_{q=u,d,s,c,b}\biggl(
1+\frac{7\alpha_s}{6\pi}
\biggr)
\{(g_V^q)^2 +(g_A^q)^2\}g_{\text{B1}}(z)
\nonumber\\
&
+ (g_V^t)^2 f_V(z,\tau) +(g_A^t)^2 f_A(z, \tau)
\biggr] ~,
\end{align}
where the functions $f_V(x,y)$ and $f_A(x,y)$ are given in
Appendix~\ref{app:massfunctions}. The twist-2 contribution is given by
\begin{align}
 C^q_{\text{T}_i}(\mu_W)=&  \frac{\alpha_2^2}{m_W^3}
\frac{n^2-(4Y^2+1)}{8}
\biggl[g_{{\rm T}_i}(w,0) \nonumber \\
&+\frac{\alpha_s(\mu_W)}{4\pi}
\biggl( - \frac{32}{9}g_{{\rm T}_i}^{\rm log}(w,0;\mu_W)
+\frac{9}{4}g_{{\rm T}_i}(w,0)
+\frac{16}{9}h_{{\rm T}_i}(w)
\biggr)
\biggr]
\nonumber \\ 
&+\frac{\alpha_2^2Y^2\{(g_V^q)^2 +(g_A^q)^2\}}{m_Z^3\cos^4\theta_W }
\biggl[g_{{\rm T}_i}(z,0) \nonumber \\
&+\frac{\alpha_s(\mu_W)}{4\pi}
\biggl( - \frac{32}{9}g_{{\rm T}_i}^{\rm log}(z,0;\mu_W)
+\frac{9}{4}g_{{\rm T}_i}(z,0)
+\frac{16}{9}h_{{\rm T}_i}(z)
\biggr)
\biggr]~,
\end{align}
for $q=u,d,s,c$, 
\begin{align}
 C^b_{\text{T}_i}(\mu_W)=&  \frac{\alpha_2^2}{m_W^3}
\frac{n^2-(4Y^2+1)}{8}
\biggl[g_{{\rm T}_i}(w,\tau)
+\frac{\alpha_s(\mu_W)}{4\pi}
\biggl( - \frac{32}{9}g_{{\rm T}_i}^{\rm log}(w,\tau;\mu_W)
\biggr)
\biggr]
\nonumber \\ 
&+\frac{\alpha_2^2Y^2\{(g_V^b)^2 +(g_A^b)^2\}}{m_Z^3\cos^4\theta_W }
\biggl[g_{{\rm T}_i}(z,0) \nonumber \\
&+\frac{\alpha_s(\mu_W)}{4\pi}
\biggl( - \frac{32}{9}g_{{\rm T}_i}^{\rm log}(z,0;\mu_W)
+\frac{9}{4}g_{{\rm T}_i}(z,0)
+\frac{16}{9}h_{{\rm T}_i}(z)
\biggr)
\biggr]~,
\end{align}
and
\begin{align}
 C^G_{{\rm T}_i}(\mu_W) =&
\frac{\alpha_2^2}{m_W^3} \frac{n^2-(4Y^2+1)}{8}
\frac{\alpha_s(\mu_W)}{4\pi}\times \nonumber \\
&
\left[4\times
\biggl(
-\frac{2}{3}g_{{\rm T}_i}^{\rm log}(w,0;\mu_W)
+\frac{1}{2}g_{{\rm T}_i}(w,0)
- \frac{2}{3}h_{{\rm T}_i}(w)
\biggr)- \frac{2}{3}g_{{\rm T}_i}^{\rm log}(w,\tau;\mu_W)
\right] \nonumber \\
+\sum_{q=u,d,s,c,b}&\frac{\alpha_2^2Y^2\{(g_V^q)^2 +(g_A^q)^2\}}{m_Z^3\cos^4\theta_W }
\frac{\alpha_s(\mu_W)}{4\pi}\biggl[
- \frac{2}{3}g_{{\rm T}_i}^{\rm log}(z,0;\mu_W)
+\frac{1}{2}g_{{\rm T}_i}(z,0)
-\frac{2}{3}h_{{\rm T}_i}(z)
\biggr]~.
\end{align}
Here we note that the LO $Z$ boson contribution to $C^q_{\text{S}}$,
$C^G_{\text{S}}$, and $C^q_{\text{T}_i}$ differs from that given in
Ref.~\cite{Hisano:2011cs} by a factor of two.

%%%%%%%%%%%%%%%%%%%%%%%%%%%%%%%%%%%%%%%%%%%%%%%%%%%%%%%%

\end{document}